\documentclass{article}

\usepackage{tcolorbox}
\usepackage{tabularx}
\usepackage{array}
\usepackage{colortbl}
\tcbuselibrary{skins}

\usepackage{algorithmic}
\usepackage{algorithm}
\usepackage{authblk}
\usepackage{booktabs}
\usepackage{amsmath}
\usepackage{graphicx}
\usepackage{comment}
\usepackage{xcolor}
\usepackage{multirow}
\usepackage{pbox}
 
\usepackage{flushend}
\usepackage{subfigure}  
\usepackage{array}
\usepackage{mathtools}

\usepackage{wrapfig}
\usepackage{graphicx}
\usepackage{balance}
\usepackage{url}

\title{Energy-Free Sensing and Context Recognition Using Photovoltaic Cells}

\author[1]{Kaede~Shintani}
\author[1,2,3]{Hamada~Rizk}
\author[1,3]{Hirozumi~Yamaguchi}
\affil[1]{\small Osaka University, Japan}
\affil[2]{\small Tanta University,  Egypt}
\affil[3]{\small RIKEN Center for Computational Science,  Japan}
\affil[ ]{\small\texttt{k-shintani@mc.net.ist.osaka-u.ac.jp, hamada\_rizk@f-eng.tanta.edu.eg, h-yamagu@ist.osaka-u.ac.jp}}

\date{}

\begin{document}

\maketitle

\begin{abstract}
The field of energy-free sensing and context recognition has recently gained significant attention as it allows operating systems without external power sources. Photovoltaic cells can convert light energy into electrical energy to power sensing devices, but their power may not be sufficient to ensure energy-free sensing due to the varying power needs of sensors and high computational demands. In this paper, we propose the use of photovoltaic cells as a standalone sensor for the recognition of different contexts, including user identification, step counting, and location tracking. The system utilizes the photocurrent readings generated by the photovoltaic cells to capture the unique mobility patterns of different users. By analyzing these patterns, the system can accurately identify the user, count the number of steps taken, and track the user's location. We propose a computationally efficient DTW to match the variable length sequences of photocurrent readings to a database of known patterns and identify the closest subject and location matches. The system was rigorously evaluated in a realistic environment, and the results indicate that it can accurately estimate step count, identify subjects, and localize them with an accuracy of 88\%, 90\%, and 43cm, respectively. This is achieved while the proposed system is non-intrusive and can operate without external power sources, making it a promising technology for energy-free sensing and context recognition. 

The code and demos are publicly available at \url{https://github.com/k-shintani/EnergyFreeStepCounting/tree/main}.

\end{abstract}

\section{Introduction}

\footnote{The code and demos are publicly available at \url{https://github.com/k-shintani/EnergyFreeStepCounting/tree/main}.}The increasing demand for wearable technology that can monitor and personalize activities is met with an array of sensor technologies, leading to the development of smart devices like watches, rings, smartphones, glasses, and necklaces that are becoming more intuitive and versatile ~\cite{s22041476,7993011}. These devices are pivotal in collecting and analyzing context information for functions like activity recognition, step counting, location tracking, and user identification.
Step counting, as an indicator of physical activity, plays a crucial role in promoting health and fitness, encouraging a more active lifestyle ~\cite{Master2022}. Moreover, wearables offering precise user identification through non-intrusive methods enhance security and user experience by tailoring preferences and linking biometric data to the correct user ~\cite{8031167}. Location detection further integrates with smart environments, enabling automated personalization like temperature control, smart lighting, and location-based content provision, thus enhancing the user's comfort and experience.

The growing dependence on mobile and wearable devices has intensified concerns about battery life, particularly with the proliferation of smart devices. Users increasingly experience "low battery anxiety,"\cite{battery_anxiety} a phenomenon substantiated by studies indicating that over 90\% of individuals suffer from this stress when their device power dwindles, especially during critical use cases like mobile payments or navigation ~\cite{battery_anxiety, tang2020alleviating}. This anxiety not only affects user comfort but also prompts behaviors like seeking charging help from strangers or reducing device usage.
To address these challenges and align with sustainability goals, the development of energy-free smart wearables is paramount.

Innovative solutions like energy harvesters (EH) are being explored to transform ambient energy into electrical power, ensuring continuous operation without traditional recharging or battery replacement ~\cite{ma2019solargest}. This shift towards self-sustaining devices not only alleviates maintenance concerns but also enhances reliability and user convenience. By negating the need for frequent battery servicing, these advancements reduce the environmental impact of battery use and support sustainable technology practices.

Several efforts have been devoted in this direction, e.g., step-counting with kinetic energy harvesters has been introduced in
 \cite{Rodriguez_2017,bhatti2016energy}. However, its harvested energy is insufficient to power the sensing and communication modules. On the other hand, 
photovoltaic (or Solar) cells are a common energy harvesting means that enjoy many advantages making them an attractive choice. A solar cell consists of semiconductor material and generates an electric current (photocurrent) in response to the ambient light energy falling on its surface \cite{sudevalayam2010energy}. This is done with relatively high conversion efficiency, better reliability, and less maintenance compared to its counterpart of energy harvesters (e.g., motion/kinetic). 
Additionally, the amount of harvested energy can encode information about the underlying physical processes. Thus, photovoltaic cells can be used as a sensor to replace conventional power-consuming sensors. This has been verified in \cite{9767256, 10.1145/3557915.3560952}, where photovoltaic cells have shown great potential for passive indoor positioning. Finally,  the market for photovoltaic cells is growing with many new innovations, such as long-life transparent cells,  which are sold at low prices, making their dependent solutions ubiquitous and cost-effective.

Motivated by these advantages, this study introduces a novel approach to wearable technology: shoes designed for net-zero energy consumption, integrating both sensing and computational capabilities. This design incorporates a compact edge device equipped with a photovoltaic cell and a computing unit integrated into the footwear. This design enables the dual functionality of energy harvesting and contextual sensing, facilitating applications such as step counting, user identification, and localization. The proposed system operates on a principle where a photovoltaic cell mounted on the shoe absorbs light and converts it into electrical energy, sufficient to power the device's computing requirements. Simultaneously, this cell serves as a sensor, where movement-induced variations in light intensity generate distinct photocurrent patterns. The analysis of these patterns for contextual recognition requires addressing several challenges, including environmental noise, variable lighting conditions, and the operation of algorithms on devices with limited resources and energy. To overcome these challenges, the system uses low-pass and differential filtering to refine the photocurrent signals and identify anomalies like missed steps.  Additionally, it employs an optimized Dynamic Time Warping (DTW) algorithm tailored for this system to accurately align the variable-length sequences of photocurrent readings with specific user movements and locations, thereby enabling precise recognition.

The system was thoroughly tested in a realistic setting with six participants and varying light conditions to mimic everyday environments. During these tests, it demonstrated high accuracy levels: 88\% in detecting steps, 90\% in identifying users, and it could track location within a 43cm median error, all with net-zero sensing and computation energy consumption. This performance highlights its potential as a leading solution in the development of sustainable wearable technologies, setting a new benchmark for energy efficiency and functionality.
Furthermore, the proposed energy-free sensing system aligns with the United Nations' Sustainable Development Goals (SDGs), particularly Goal 7: Affordable and Clean Energy~\cite{GIELEN201938}, Goal 9: Industry, Innovation, and Infrastructure~\cite{Singh2023}, and Goal 11: Sustainable Cities and Communities~\cite{doi:10.1177/0956247815627522}. By utilizing photovoltaic cells for both sensing and energy harvesting, the system promotes energy efficiency and sustainability, reducing dependence on batteries and external power sources. This directly supports SDG 7 by fostering the adoption of renewable energy technologies. Additionally, the system's application in human activity recognition and localization enhances smart infrastructure development (SDG 9) and contributes to more efficient and sustainable urban mobility solutions (SDG 11). Such innovations play a crucial role in the advancement of self-powered wearable technologies, ensuring both environmental and societal benefits.

\section{Motivation and Background}
\subsection{Motivation for Solar Cells}
\textbf{Efficiency: }
Solar cells, also known as photovoltaic cells, are a widely used and cost-effective source of energy for powering the Internet of Things (IoT) sensor nodes~\cite{sandhu2021solar,ma2019sensing}. These cells consist of a semiconductor material that generates an electric current in response to ambient light, making them suitable for both indoor and outdoor environments. This makes them useful for a variety of applications, including handheld calculators, garden lights, and wearable devices~\cite{ma2018gesture,ma2019solargest,ma2019sensing}.
Solar cells offer advantages in terms of conversion efficiency and robustness over other forms of energy harvesting such as kinetic energy. High conversion efficiency allows for a greater proportion of energy to be extracted from the source, while robustness means that the system is reliable, requires minimal maintenance, and consistently produces a similar output when exposed to similar environments. The amount of energy harvested by a solar cell depends on the intensity of light and the orientation of the cell relative to the light source. When worn on the human body, the harvested energy changes as a result of different mobility patterns relative to the light source, which can be used to recognize different contexts such as identifying the wearer, following her steps, and tracking her location.

\textbf{Proliferation and Market: } The global market for indoor photovoltaic (IPV) cells reached \$140 million USD in 2017 and is expected to experience significant growth in the coming years~\cite{MATHEWS20191415}. Market studies predict that the annual IPV market size will reach \$850 million USD by 2023 and \$1 billion USD by 2024. This growth is driven by the increased demand for photovoltaic cells, with an estimated 60 million devices to be sold annually. This market is growing at a much faster pace compared to other energy harvesting technologies~\cite{AMR2021}.
Different vendors are taking advantage of this growth by incorporating IPV cells into their products for energy harvesting. Additionally, the development of transparent and thin IPV cells presents new opportunities for context-aware sensing and energy harvesting in indoor and outdoor environments. This is a particularly unique opportunity for enabling battery-less sensing for context-aware extraction of human-centric applications.
Recent studies have also explored the integration of semitransparent thermoelectric cells (TECs) for energy harvesting, which not only generate electricity but also allow light transmission and heat water~\cite{Karimov2020}. Furthermore, advancements in transparent photovoltaic (TPV) technology have enabled its integration into self-powered memory devices, demonstrating their potential for energy-efficient storage applications~\cite{10.1002/solr.202300720}. These TPVs can facilitate the development of step-count-based memory elements, where photovoltaic-generated power supports real-time gait data storage, eliminating the need for external batteries.
\color{black}

\subsection{Background}
Dynamic Time Warping (DTW) ~\cite{Mller2007} is a technique for measuring the similarity between two temporal sequences. It is commonly used in speech recognition, time series forecasting, and other applications where the alignment of time series data is important.

Let $X={x_1,x_2, \dots, x_n}$ and $Y={y_1,y_2, \dots, y_m}$ be two sequences of equal or different lengths. The goal of DTW is to find an optimal alignment between the two sequences, such that the accumulated distance between the aligned elements is minimized. The accumulated distance between two aligned elements $x_i$ and $y_j$ is given by a distance function $d(x_i,y_j)$.

DTW can be defined mathematically using a matrix $D$ where $D_{i,j}$ is the accumulated distance between the first i elements of $X$ and the first j elements of $Y$. The last element $D_{n,m}$ gives the optimal alignment distance between the two sequences~\cite{Mller2007}.

\begin{equation}
D_{i, j} = d(x_i,y_j) + min(D_{i-1, j}, D_{i, j-1}, D_{i-1, j-1})
\end{equation} 

The DTW-based similarity measure between two sequences is defined as the minimal accumulated distance between the two sequences over all possible alignments.
\begin{equation}
    sim(X,Y) = D_{n,m}
\end{equation}

DTW can be thought of as a flexible method~\cite{pmid31523672, sym13050836} for comparing sequences that can deal with variable-length, non-linear stretches and shifts between the two sequences being compared. The main idea is to use a sliding window to iteratively align the two sequences at different points in time, with the goal of minimizing the accumulated distance between the aligned elements. Thus DTW is used as a similarity measure for classifying time series with methods like k-NN.

\begin{figure}[!t]
\begin{center}
\includegraphics[width=1\linewidth]{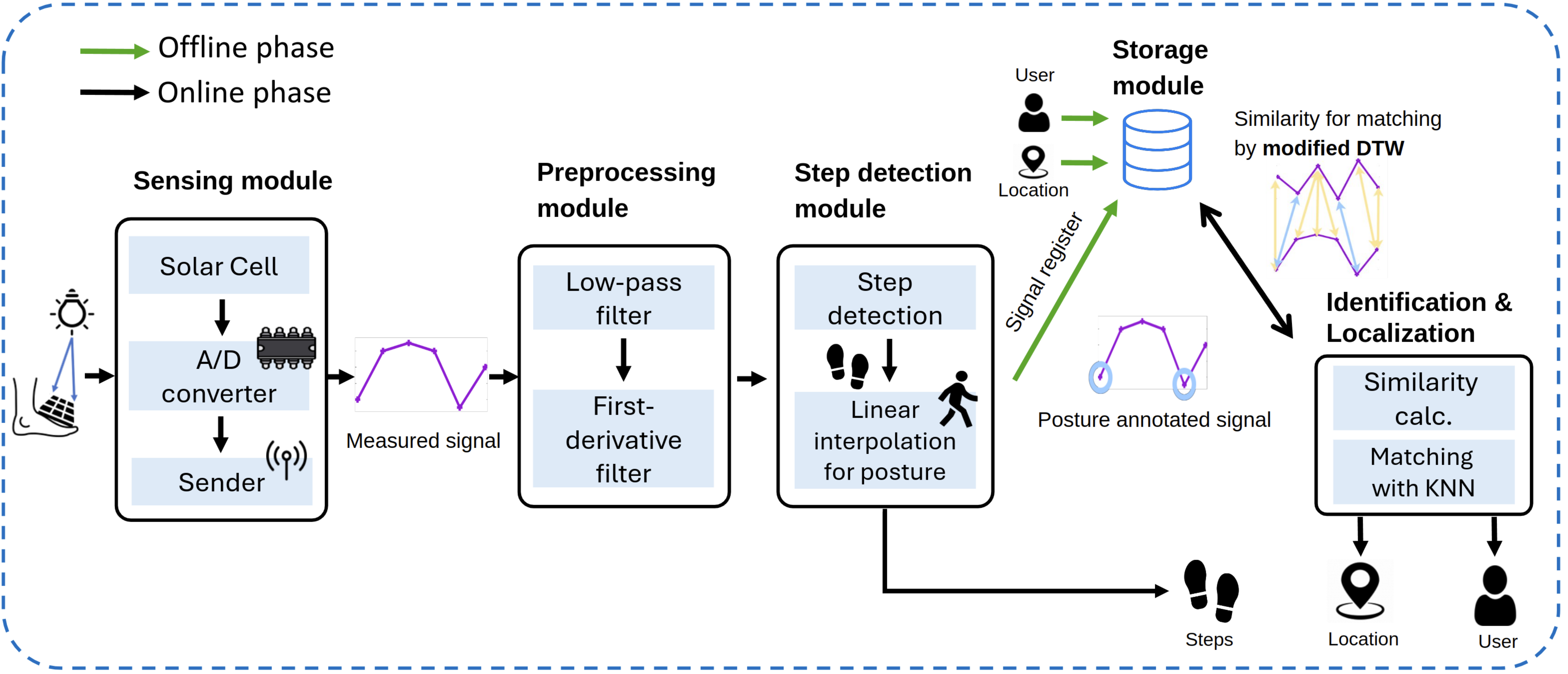}
\caption{The Proposed system architecture.}
\label{fig:system_architecture_real_time}
\end{center}
\end{figure}

\section{System Overview}
The proposed system consists of two phases: an offline phase and an online phase, as shown in Figure~\ref{fig:system_architecture_real_time}. The offline phase is composed of the Sensing and Data Collection module, which is responsible for capturing photocurrent measurements via a compact-sized device mounted on the user's shoes. The measurements are taken while the user walks naturally in the area of interest. These measurements, along with the user's identification and reference location, are recorded using a data collection app installed on a mobile device. Subsequently, the collected data is sent to a cloud-based service to construct a fingerprint database. The Preprocessing module is then utilized to process the data, which includes the mitigation of noise, cleaning of the data, and smoothing of the measurements. The purpose of these preprocessing steps is to facilitate context recognition in the online phase.

The online phase is made up of three primary modules: Step Detection, Human Identification, and Location Estimation. The captured photocurrent measurements on the run time are utilized to detect the user's steps and foot posture through the use of differential filters and peak recognition. Then, the estimated step information along with the fingerprints gathered during the offline phase information are leveraged to identify the user by analyzing her mobility pattern. Additionally, the Location Estimation module uses the captured photocurrent to match with the recorded fingerprint of various reference locations to determine the current user location.

\begin{figure}[!t]
    \centering
    \subfigure[Photovoltaic cell mounted on a shoe.]{
        \includegraphics[width=.42\linewidth,height=3.cm]{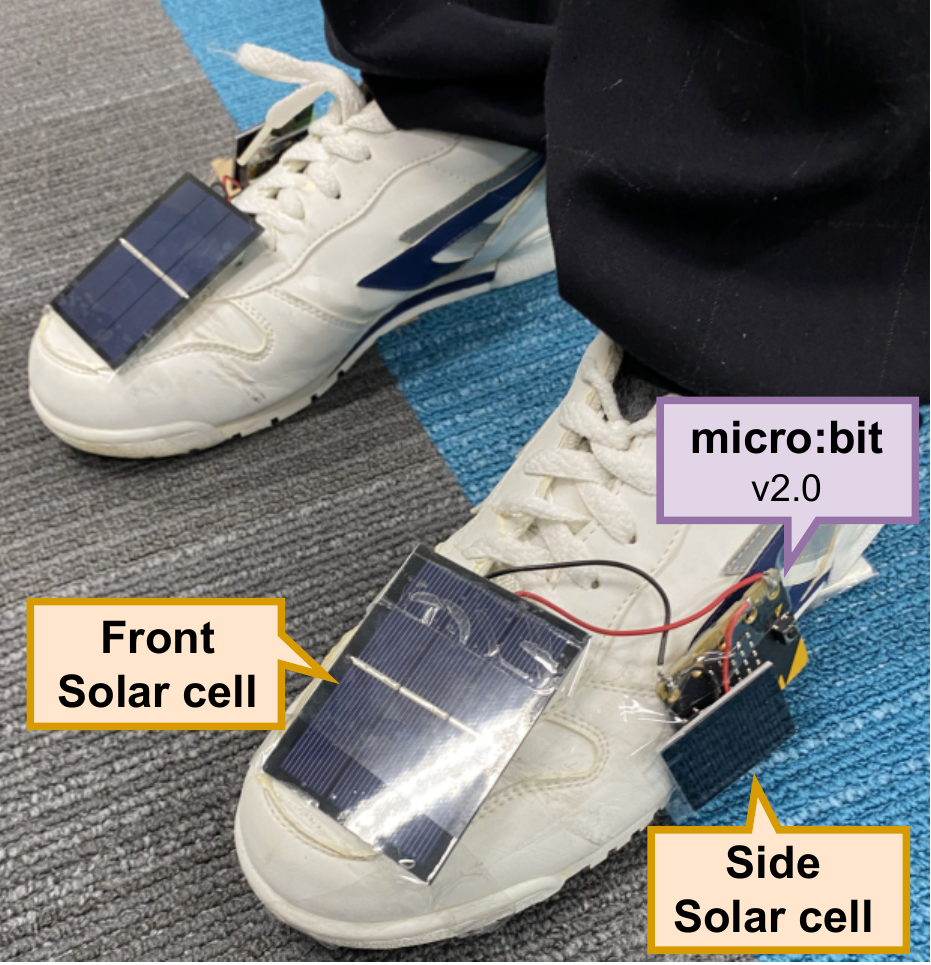}
        \label{fig:demo_shoes}
    }
    \subfigure[The circuit design.]{
        \includegraphics[width=.5\linewidth,height=3.cm]{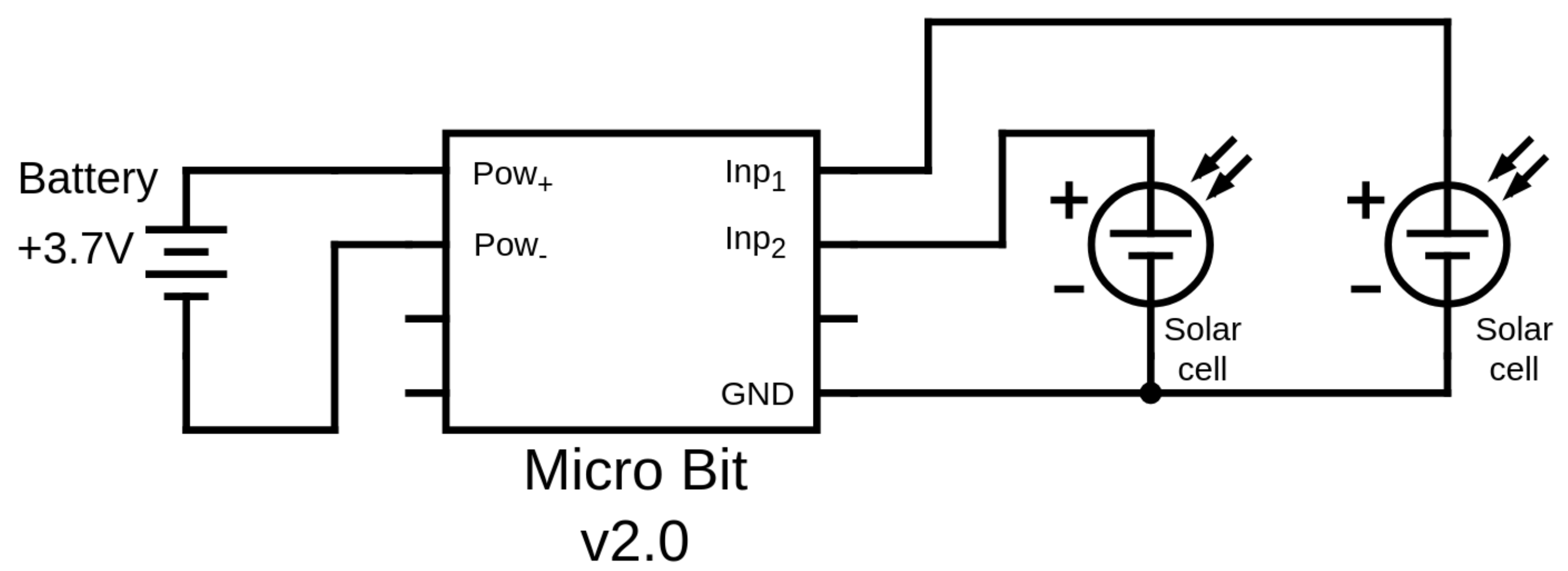}
        \label{fig:real_setup_cir}
    }
    \caption{Shoes-mounted photovoltaic cell and illustration the connection of various components.}
    \label{fig:experiment}
   
\end{figure}

\section{The System Details} 
Figure~\ref{fig:system_architecture_real_time} shows the architecture of the proposed system in real-time.
The system involves two main parts, the sensing part and the processing part.

\begin{figure}[!tbp]
\centering
    \includegraphics[width=0.85\linewidth,height=3.8cm]{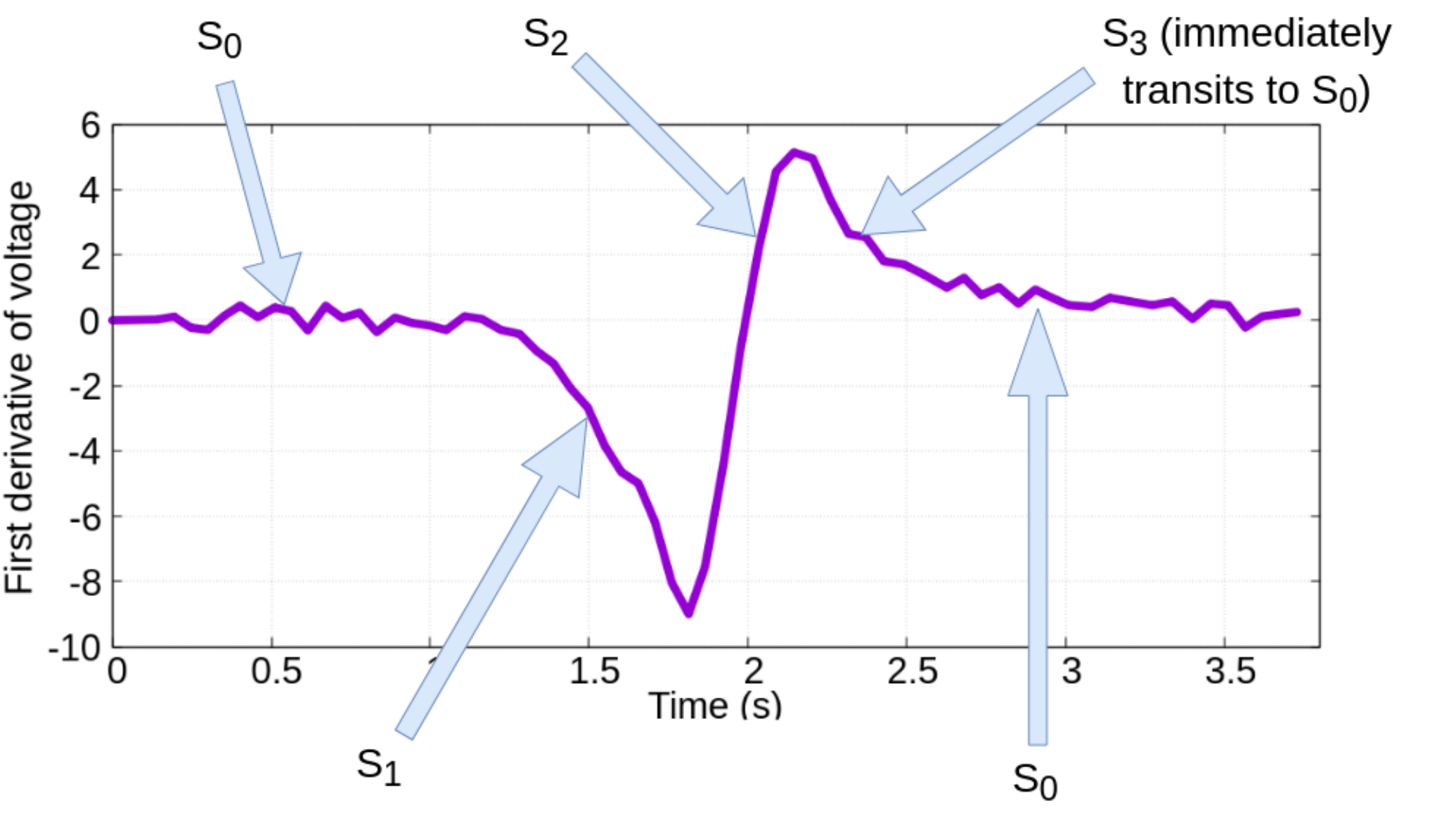}
    \caption{ Step signature. 
}
    \label{fig:states_in_single_step}
\end{figure}

\subsection{The Sensing Module}
The Sensing module of the proposed system consists of   three main components mounted to the smart shoes (see Fig.~\ref{fig:demo_shoes}): the Solar cell, the electronic circuit, and the computing unit.
Specifically, this module utilizes a commodity solar cell with dimensions of 8cm in length, 6cm in width, and 0.1cm in depth. The solar cell has an efficiency of 15.5\%. To obtain photocurrent measurements from the solar cell, it is connected to an analog-to-digital converter (ADC), which is 12-bit MCP3204. This ADC is then connected to a low-energy-compute and transmission unit, specifically a Microbit, through a Serial Peripheral Interface (SPI). The measurements are obtained using a Python implementation, which sends an HTTP request to the Microbit to get the photocurrent reading response from the solar cell. 
The photocurrent readings are further converted into voltage for ease of interpretation 
by measuring the open circuit voltage, which depends on luminous intensity.
The captured voltage measurements are then forwarded to the processing module for further analysis.

\subsection{The Preprocessing module}  \label{sec:processing_module}
This module is responsible for refining the collected voltage measurements to enhance step detection accuracy. 
The module begins with a synchronization step that ensures consistent acquisition of light intensity data from sensors attached to both feet, facilitating simultaneous analysis.
The procedure commences with the separate sampling of digital data from each foot sensor, noting that the temporal alignment of these data streams may vary. To address this, linear interpolation is employed between two consecutive light intensity readings, aligning them along the time axis, thus standardizing the temporal framework for subsequent analyses.

Signal smoothing is a critical process within this module, aimed at mitigating distortion or noise that might mask the essential step pattern. This is achieved through the application of two specific filters: a low-pass filter and a differential filter. The low-pass filter plays a pivotal role in removing high-frequency noise, thereby smoothing the signal. The mathematical representation of the smoothing process conducted by the low-pass filter is articulated as follows~\cite{Proakis2013}:

\begin{equation}
V_i = \alpha V_{i-1} + (1 - \alpha) v_i
\end{equation} 

where $v_i$ denotes the $i^{t h}$ raw voltage reading within a specified window, $V_i$ represents the $i^{t h}$ smoothed value post-filter application, and $\alpha$ is a factor controlling the influence of previous window's data.

In an endeavor to fortify the system’s resilience against fluctuations in ambient lighting conditions, the module incorporates a first-derivative filter. This filter accentuates the relative voltage change, thereby facilitating a more nuanced and accurate identification of step patterns, especially in environments with variable lighting. The first-derivative filter mathematically quantifies the instantaneous rate of voltage change as follows \cite{Shukla2014}:

\begin{equation}
V^{\prime} = \frac{V_i - V_{i-1}}{\Delta t}
\end{equation}

where $V^{\prime}$ signifies the derived rate of change in the smoothed voltage, $V_i$ is the $i^{t h}$ processed voltage value, and $\Delta t$ represents the temporal interval between successive measurements.

This differential approach is particularly adept at highlighting the cyclical nature inherent in pedestrian locomotion, effectively transforming the voltage signal into a representative waveform, thereby elucidating the step dynamics as depicted in Figure \ref{fig:states_in_single_step}.

\begin{figure}[t]
\begin{center}
\includegraphics[width=0.95\linewidth,height=4cm]{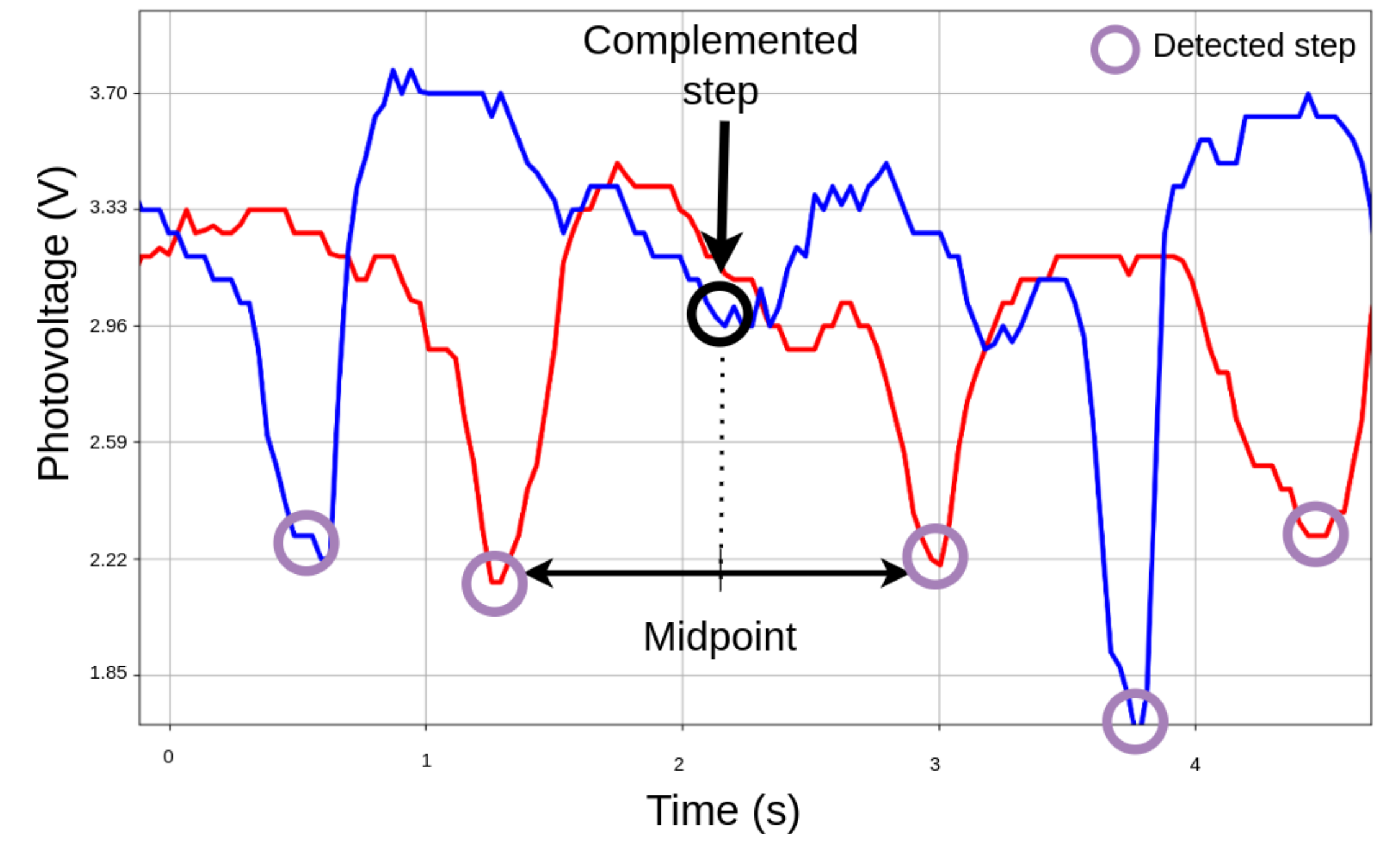}
\caption{ complemented step.}
\label{fig:complement}
\end{center}
\end{figure}

\begin{figure}[!t]
\begin{algorithm}[H]
    \caption{Missing step complement}
    \label{alg:step_comp}
    \begin{algorithmic}
    \REQUIRE{$stepDetected$ : "True" only at the moment the step detection module detects a step.}
    \REQUIRE{$foot_{prev}$ : which foot was used to detect the step in the last}
    \REQUIRE{$foot_{now}$ : which foot is
currently used to detect step}
    \REQUIRE{$t_{prev}$ : last step detection time}
    \REQUIRE{$t_{now}$ : the current step detection time}
    \REQUIRE{$t_{mid}$ : the newly completed mean}    
    \REQUIRE{$t_{THR}$ : thresholds for the maximum allowable time difference when complementing}
    \ENSURE{$detectedSteps$ : the time of the step occurrence and the right or left foot}
    \IF{$stepDetected$}
        \IF{$foot_{prev} = foot_{now}$}
            \IF {$t_{now} - t_{prev} < t_{THR}$}
                \STATE { $t_{mid} \gets (t_{now} + t_{prev}) / 2$}
                \STATE { $foot_{opposite} \gets opposite( foot_{now} ) $}
                \STATE { $detectedSteps.add(t_{mid}, foot_{opposite})$}
            \ENDIF
        \ENDIF
        \STATE  {$detectedSteps.add(t_{now}, foot_{now}$)} 
        \STATE  {$t_{prev} \gets t_{now}$} 
        \STATE  {$foot_{prev} \gets foot_{now}$} 
    \ENDIF 
    \end{algorithmic}
\end{algorithm}
\end{figure}

\subsection{The Step Detection Module}
The Step Detection Module plays a pivotal role in monitoring and interpreting the dynamic electrical signatures generated by foot movements during locomotion, specifically through the analysis of voltage amplitude variations. These variations follow a distinctive pattern associated with the stepping action, characterized by an initial voltage level, a subsequent decline, and a notable increase, indicative of the foot's varying exposure to the light source due to changes in orientation or position. This cyclical voltage pattern, indicative of the step cycle, is shown in Figure \ref{fig:states_in_single_step}.
To effectively capture and analyze these patterns in real-time, this module employs a peak detection algorithm. This algorithm is particularly advantageous for its low computational demand, making it ideal for edge computing environments where resources are limited. It monitors the voltage trajectory, identifying peaks that correspond to the step events, and thereby facilitates the initiation of the step counting process upon recognition of the characteristic step pattern.

Moreover, the module is designed to recognize and rectify anomalies in the step sequence, such as consecutive steps from the same foot within a short interval, typically less than 2 seconds. In such instances, it infers and inserts a complementary step for the opposite foot at the mid-point of the interval, thereby enhancing the accuracy of step counting as detailed in Algorithm~\ref{alg:step_comp}.

\begin{figure}[t]
\begin{center}
\includegraphics[width=1\linewidth]{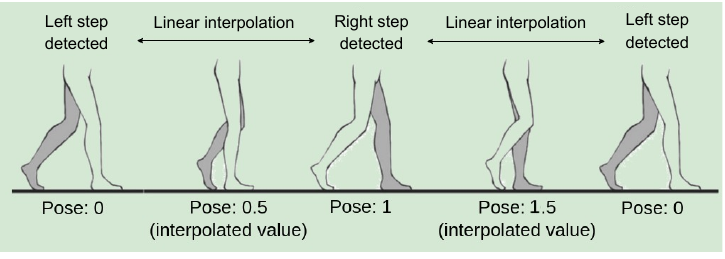}
\caption{Cyclic pattern of the foot posture.}
\label{fig:posture}
\end{center}
\end{figure}

\color{black}
The system employs a \textit{Linear Interpolation process} for dynamically quantifying foot posture (detailed in Algorithm~\ref{alg:interpolation}.
), thereby enhancing user identification and positioning accuracy. This process is predicated on the notion that each individual's walking pattern is distinct, with unique foot extension and orientation. By assigning a numerical scale from 0 to 2, where 1 represents the full extension of the left foot and 0 the right foot's full extension, the system can capture the entire gait cycle's nuances. This linear interpolation translates biomechanical movements into a quantifiable data stream, mirroring the natural, cyclic progression of walking, as shown in Figure~\ref{fig:posture}. It effectively captures the transitions between peak foot extensions, creating a detailed and continuous profile of foot motion that serves as a unique signature for each user, thus aiding in precise user matching and identification.

\begin{figure}[t]
\begin{center}
\includegraphics[width=1\linewidth]{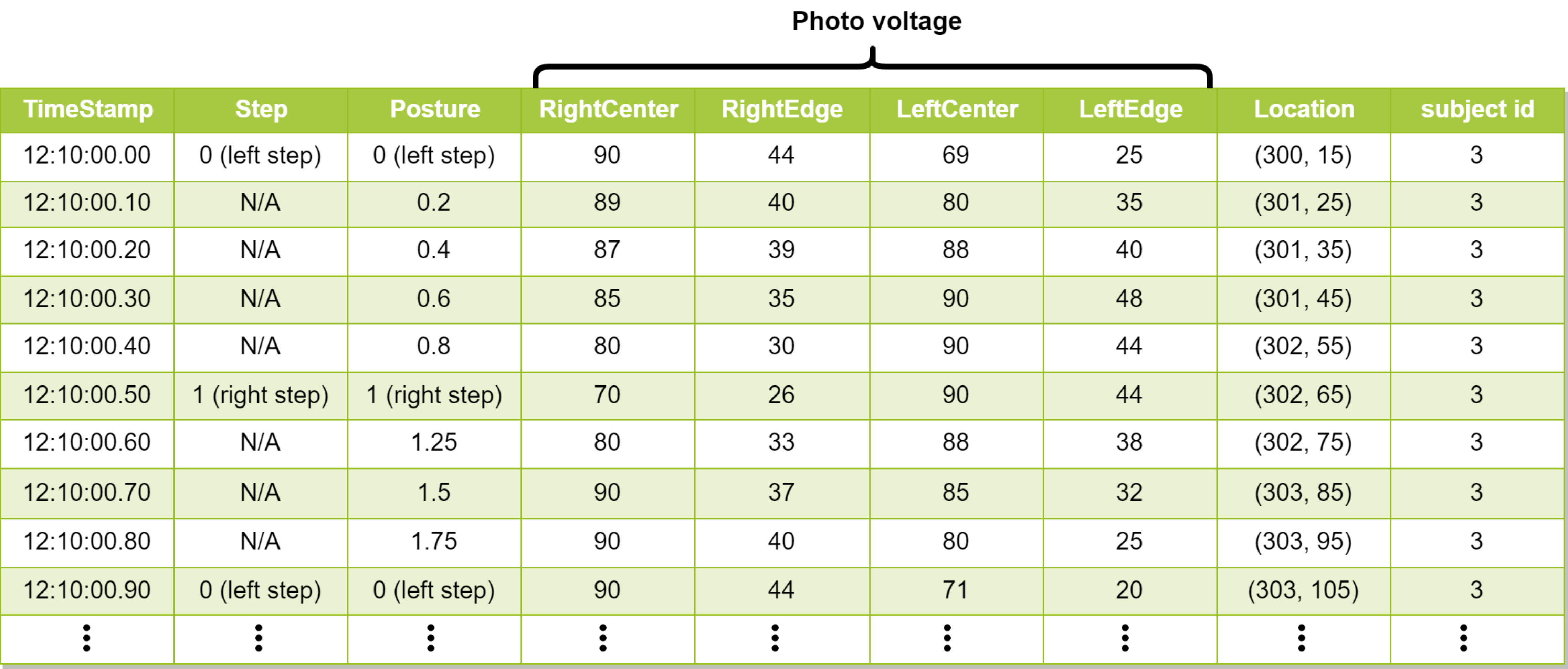}
\caption{  Stored features.}
\label{fig:features}
\end{center}
\end{figure}

Complementing the Linear Interpolation process, the \textit{Storage Module} forms a crucial component of the system’s architecture, dedicated to the aggregation of pertinent time-series data that encapsulates crucial aspects of the stepping process. The dataset comprises timestamps, detected steps, interpolated foot postures, and variations in photovoltaic voltage from the solar cell sensors and is augmented with location coordinates and the subject’s identification labels, as exemplified in Figure \ref{fig:features}.

Together, these elements provide a holistic and integrated approach to monitoring and analyzing gait dynamics, facilitating a deep understanding of individual walking patterns. This data enables the system to perform the user identification and localization tasks, as described in the following section.

\begin{figure}[!t]
\begin{algorithm}[H]
    \caption{Linear interpolation}
    \label{alg:interpolation}
    \begin{algorithmic}
    \REQUIRE{$stepDetected$ : "True" only at the moment the step detection module detects a step}
    \REQUIRE{$foot_{prev}$ : which foot was used
to detect the step in the last}
    \REQUIRE{$foot_{now}$ : which foot is currently used to detect step}
    \REQUIRE{$foot_{inter}$ : interpolated posture value between current detected step and the last step}
    \REQUIRE{$t_{prev}$ : last step detection time}
    \REQUIRE{$t_{now}$ : the current step detection time}
    \REQUIRE{$t_{interval}$ : the interval time between interpolated data}
    \ENSURE{$interpolateds$ : sequences of the time of the posture occurrence and the interpolated posture value }
    \IF{$stepDetected$}
        \STATE { $t \gets t_{prev}$}
        \WHILE{ $t < t_{now}$}
            \STATE {$foot_{inter} \gets foot_{prev}+(t-t_{prev}) * abs(foot_{now}-foot_{prev}) / (t_{now}-t_{prev})$}
            \STATE { $interpolateds.add(t, foot_{inter}) $}
            \STATE {$t \gets t + t_{interval}$}
        \ENDWHILE
        \STATE  {$t_{prev} \gets t_{now}$} 
        \STATE  {$foot_{prev} \gets foot_{now}$} 
    \ENDIF 
    \end{algorithmic}
\end{algorithm}
\end{figure}

\begin{figure}[t]
\begin{center}
\includegraphics[width=.95\linewidth,height=4.2cm]{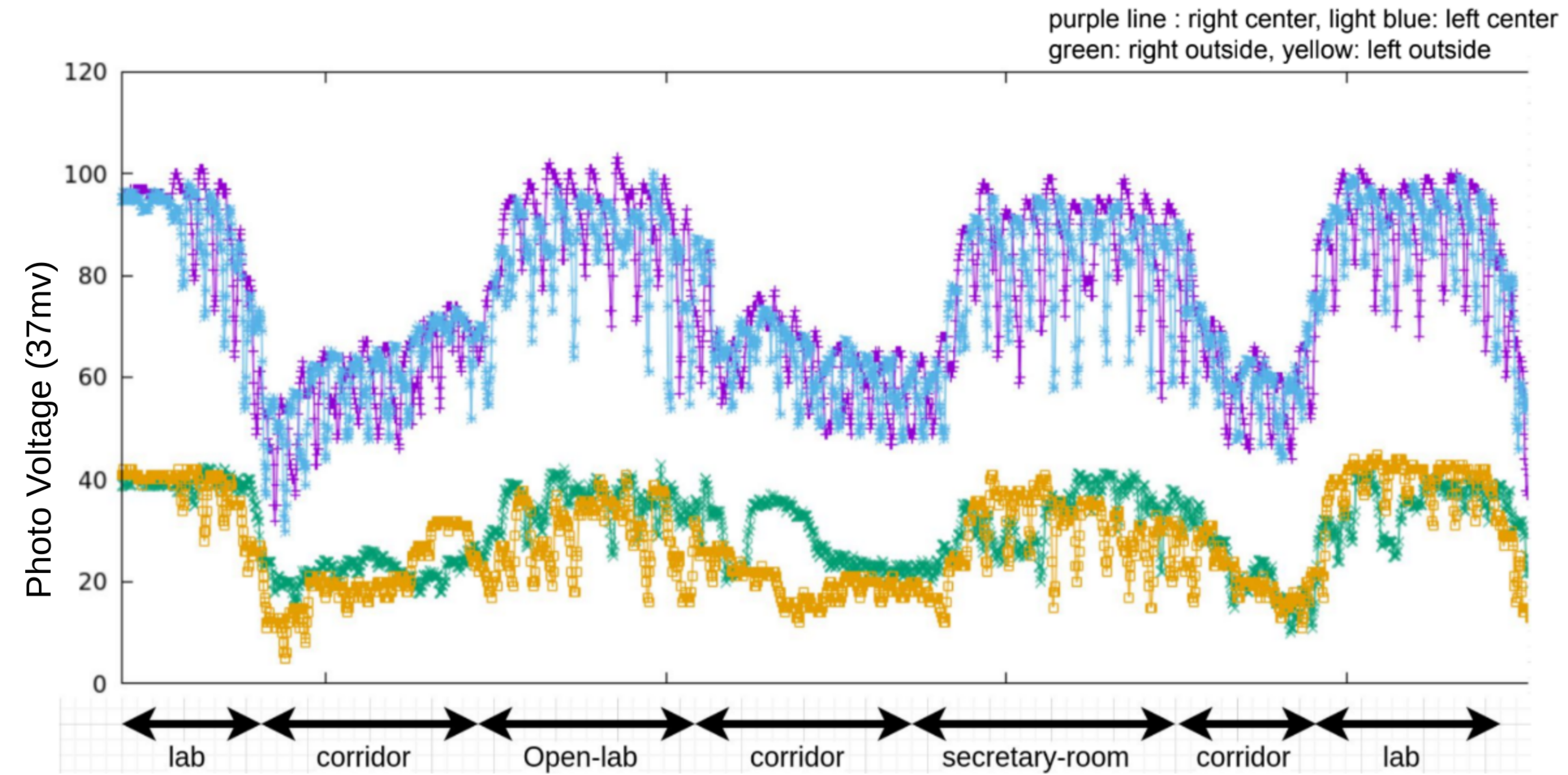}
\caption{Photovoltage Measurements Across Various Rooms: This graph demonstrates the capacity to distinguish between different locations based on variations in photovoltage readings.}
\label{fig:loc_photocurrent}
\end{center}
\end{figure}

\subsection{Subject Identification and Localization}
Our proposed system operates in an online mode, designed specifically to match observed mobility patterns against a pre-stored database for two primary tasks: Subject Identification and Location Estimation. By capturing and analyzing unique mobility patterns through changes in photocurrent readings and foot posture, the system is able to identify individuals and determine their location with high precision.

For Subject Identification, the system utilizes the k-nearest neighbors (KNN) algorithm to compare an observed sequence of mobility data over a 5-second interval against known patterns of different users stored in the database. Similarly, for Location Estimation, the system compares the observed data against stored patterns of different locations (as depicted in Fig~\ref{fig:loc_photocurrent}). This dual functionality enables effective monitoring and tracking within the operational environment.

A critical aspect of using KNN effectively involves choosing an appropriate distance metric for time-series data. Commonly used Euclidean distance may not be suitable due to variations in step length over time. Therefore, we employ the Dynamic Time Warping (DTW) algorithm, which adjusts for data stretching and shrinking along the time axis. Standard DTW, while robust, is computationally expensive with a complexity of $O(|S| \times|T|)$~\cite{Mller2007}, where $S$ and $T$ are the sequence lengths. To overcome this, we introduce a Modified DTW approach (described in Algorithm~\ref{alg:DTW}) that reduces computational demand to a linear complexity of $O(|S|+|T|)$ without compromising accuracy. This is possible by utilizing previously estimated foot postures from another module to enhance the matching process.

\begin{figure}[H]
\begin{center}
\includegraphics[width=1\linewidth]{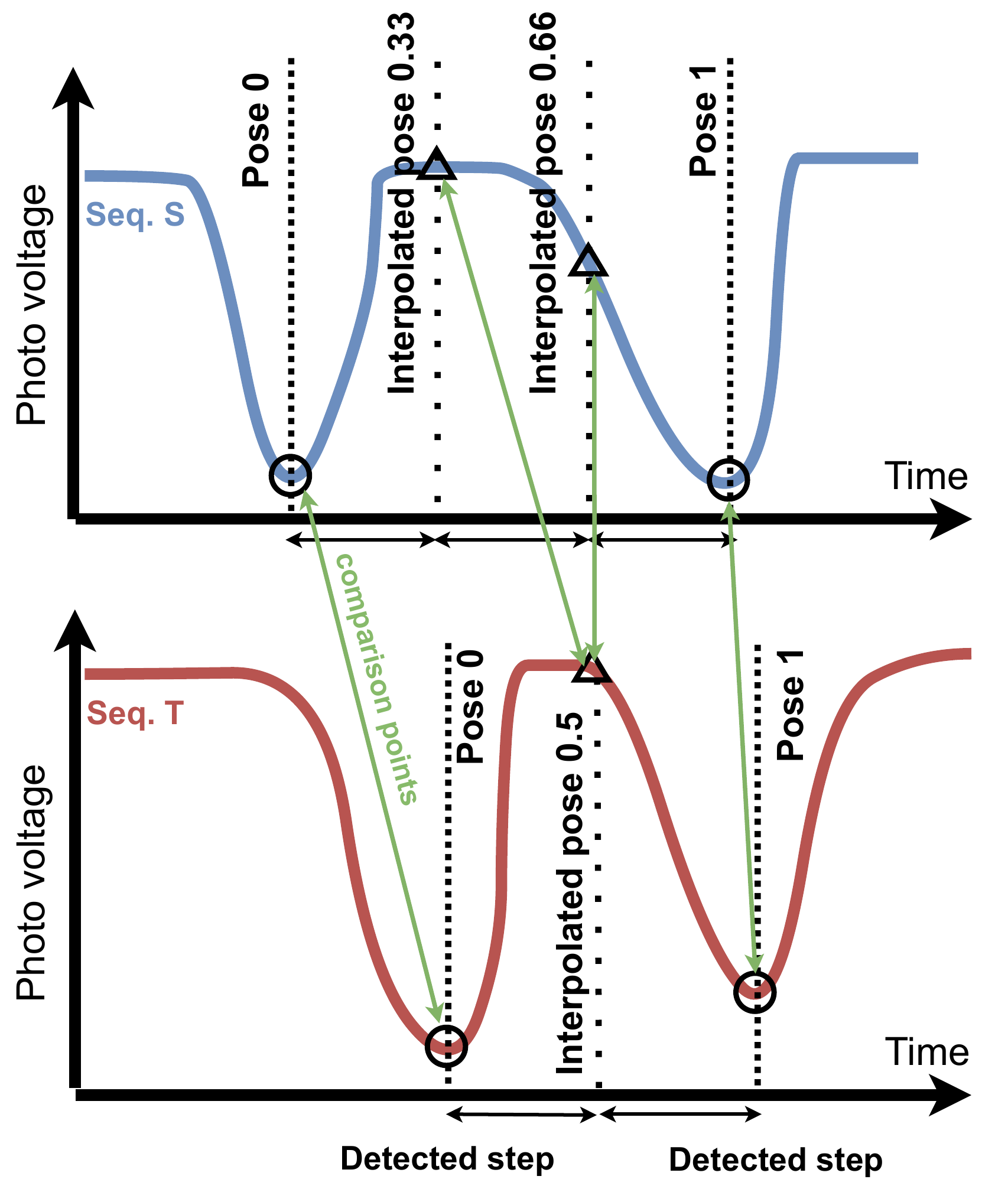}
\caption{Example on how the Modified DTW algorithm works.}
\label{fig:dtw}
\end{center}
\end{figure}

The sequences $S=s_1, s_2, \ldots, s_{|S|}$ and $T=t_1, t_2, \ldots, t_{|T|}$ represent time-series data of photovoltage measurements. Foot posture phases for each point in these sequences are determined using a function $p()$, which identifies user steps with a step detector module and applies linear interpolation to estimate foot posture at each step. Our Modified DTW constructs a warp path $W=w_1, w_2, \ldots, w_K$ that effectively aligns $S$ and $T$, matching similar elements while maintaining temporal order. This warp path starts at $w_1=(1,1)$ and ends at $w_K=$ $(|S|,|T|)$, covering the entire length of both sequences. It adheres to constraints ensuring it progresses in a non-decreasing order and avoids any element skipping, thus ensuring a smooth and continuous alignment. The optimization goal is to minimize the cumulative distance along this path, calculated as~\cite{Mller2007}:
$$
T W D=\sum_{k=1}^K \operatorname{Dist}\left(s_{w_{k i}}, t_{w_{k j}}\right)
$$
where $\operatorname{Dist}\left(s_i, t_j\right)$ is the Euclidean distance between elements. The flexibility of the warp path allows it to stretch in three possible directions-horizontal $(i, j+1)$, vertical $(i+1, j)$, or diagonal $(i+1, j+1)$-to accommodate variations in movement rates between the sequences.
This adaptation ensures alignment of similar movements even if they occur at different times, thereby efficiently handling time distortions, as shown in Fig.~\ref{fig:dtw}. The selective comparison of elements, informed by additional data such as foot posture phases, reduces the number of plausible paths, thus dramatically reducing the computational complexity from quadratic to linear. This strategic optimization not only ensures precise temporal alignment but also supports real-time application requirements, making this system ideal for scenarios requiring rapid and accurate subject identification based on distinct mobility patterns.

\begin{figure}[!t]
\begin{algorithm}[H]
    \caption{Modified DTW distance}
    \label{alg:DTW}
    \begin{algorithmic}[1]
     \REQUIRE {
        $s$ $ \AND $ $t$ : multi-dimensional time series data to be compared
    }
    \REQUIRE {$ p(s_i)$ $ \AND $ $ p(t_j)$ : interpolated posture values for signals s and t, referenced by $DistPosture(i, j)$}
    \REQUIRE {$Weight$ : Allowable parameter for time warping.}
    \REQUIRE {$THR_{prune}$ : Threshold for branch and bound based on large posture differences between two data heads.}
    \REQUIRE {$selectMin$ : Function that selects the smallest
of the three arguments and returns the tuple of indices $i$ and $j$ with 1 added if not warping.}

    \ENSURE {$DTWdist$ : Similarity with the time warp between two time series data, $s$ and $t$}

    \STATE {$THR_{prune} \gets 0.1$}
    \IF{$DistPosture(0, 0) > THR_{prune}$}
        \RETURN {$Inf$}
    \ENDIF
    \STATE {$i \gets 0$}
    \STATE {$j \gets 0$}
    \STATE {$DTWdist \gets 0$}
    \WHILE{ $i + 1 < |S| $ $ \AND $ $ j + 1 < |T| $}
        \STATE {$DTWdist $ $ \gets $ $DTWdist + EucDist(s[i] - t[j])$}
        \STATE {$selected \gets selectMin( $}
        \STATE {$\quad DistPosture(i + 1, j + 1),$}
        \STATE {$\quad DistPosture(i + 1,  j)  ,$}
        \STATE {$\quad  DistPosture(i,  j + 1) ,$}
        \STATE {$)$}
        \IF{$selected = (i+1, j+1)$}
            \STATE{$i \gets i + 1$}
            \STATE{$j \gets j + 1$}
        \ELSIF{$selected = (i+1, j)$}
            \STATE{$i \gets i +  1$}
        \ELSIF{$selected = (i, j+1)$}
            \STATE{$j \gets j + 1$}
        \ENDIF
    \ENDWHILE
    \RETURN {$DTWdist$}
    \end{algorithmic}
\end{algorithm}
\end{figure}

\section{Design Issues and Solution}
\color{black}

One of the issues of our system design is the relatively high energy consumption of the computation and communication on edge. 
For the computational part of step counting on the edge, processing data on the edge devices is typically performed by microcontrollers (MCUs) or single-board computers, which may not have sufficient power to operate in indoor dim light conditions. 
This can be tackled with either using a more energy efficient chip like QN9080 or using logic circuit. The QN9080 includes a MCU, on-chip memory,  Bluetooth Low-Energy radio, 16-bit ADC and low power sleep timer. These features help to run the chip for our application-specific process with very low power consumption, as quantified in Section~\ref{sec:eval_energy_consumtion}.

\begin{figure}[t]
\begin{center}
\includegraphics[width=.95\linewidth,height=4.2cm]{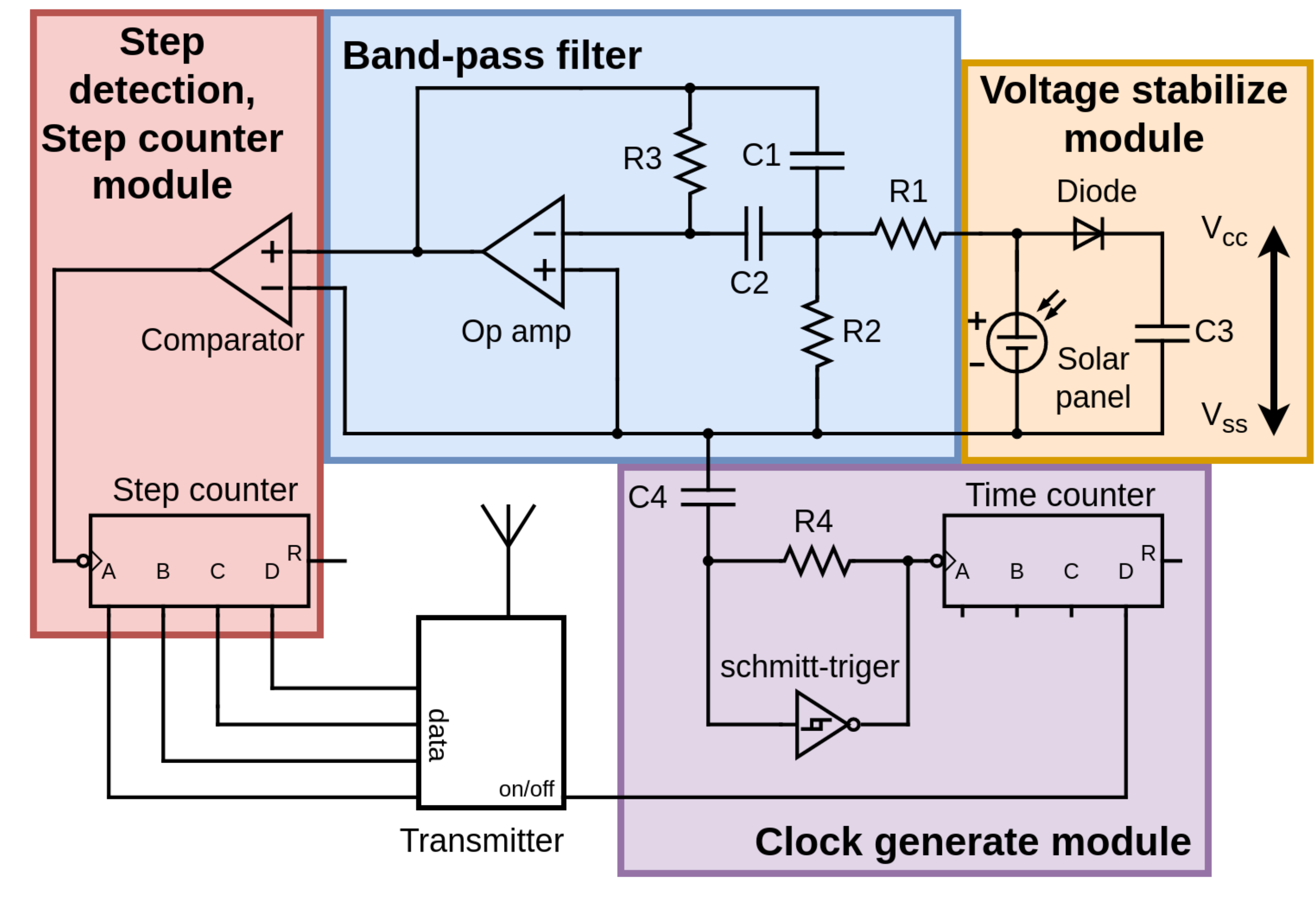}
\caption{ The circuit architecture.
}
\label{fig:circuit}
\end{center}
\end{figure}

The second solution to this problem is to replace the MCU or single-board computer with a logic circuit to operate an ultra-low power step detection circuit.
The design of this circuit is shown in Figure~\ref{fig:circuit} and is made up of a few basic components, including a \textit{voltage stabilize}, a \textit{clock generator}, and a \textit{smoothing module}. 
The voltage stabilizer module is used to stabilize the varying power generated by the solar panels using a capacitor or a small battery. Additionally, a chopper is used to boost the voltage to a level that allows the entire circuit to operate properly.
The clock generator module  generates a clock for the counter module using a Schmitt-triggered oscillation circuit.
The smoothing module is used to extract the features from the photocurrent, enabling the recognition of human steps. This module is equivalent to low-pass and differential filters introduced in Section~\ref{sec:processing_module}. 
Finally, the step detection module is responsible for converting the analog wave input into a digital wave of 0 or 1 using a Schmitt trigger with a predetermined threshold value. This module detects the step waves and sends them to the step counter module.
The step counter module is a basic flip-flop circuit that counts the step waves and stores the number of steps. 

On the other hand, the energy used for wireless communication is generally the bottleneck of the entire system. 
One potential solution to this problem is to replace the Bluetooth Low Energy (BLE) transmitter with a backscatter system. Backscatter communication is a technique that allows devices to communicate wirelessly by reflecting existing RF signals, rather than generating their own RF signals. This approach can significantly reduce the energy consumption of the communication, as the device only needs to modulate the reflected signal, rather than generating its own signal.

\textit{In summary, the proposed system is designed to be energy-efficient using a logic circuit to operate ultra-low power step detection computing and communication circuits.}

\begin{figure}[t]
\begin{center}
\includegraphics[width=0.8\linewidth]{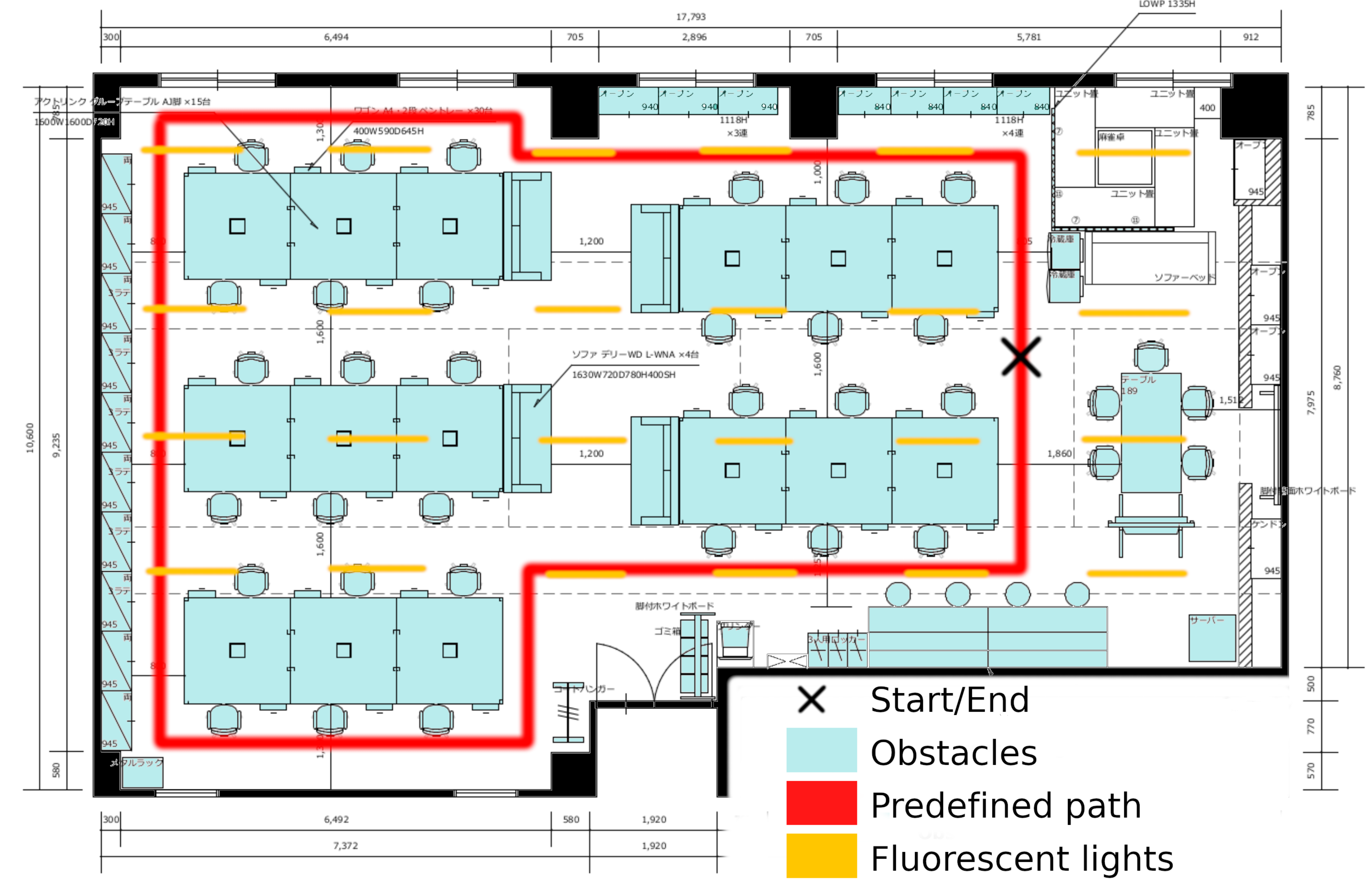}
\caption{The floorplan of the testbed.}
\label{fig:testbed}
\end{center}
\end{figure}

\begin{figure}[H]
\begin{center}
\includegraphics[width=0.6\linewidth]{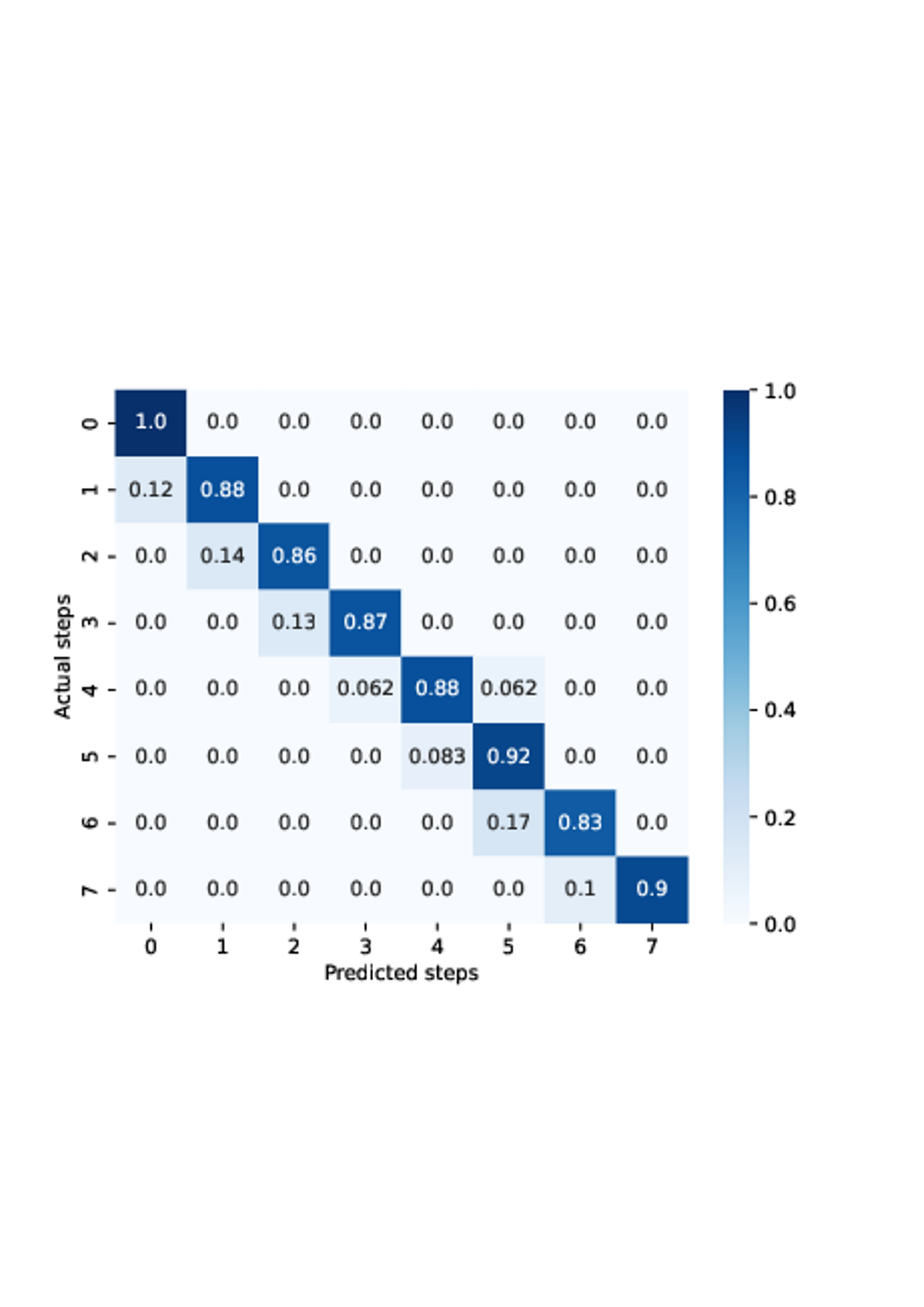}
\caption{
   { Confusion Matrix for Step Count Estimation: This matrix illustrates the accuracy of the step detector in estimating the correct step count as the user walks varying numbers of steps.}
}
\label{fig:confusion_matrix}
\end{center}
\end{figure}

\vspace{5pt} 
\begin{table}[h]
\centering
\caption{Relative / Abs error rate of step detector per step.}
\label{Tab:step_detector_error}

\begin{tabular}{c|c} 
\hline
Location & Error rate $\pm$ Standard deviation \\ 
\hline
Inside Lab (Fig.\ref{fig:testbed})  
& -0.901\% ± 0.903\% / 0.951\% ± 0.850\%  \\ 
\hline
Room to room  
& -5.82\% ± 2.91\% / 5.89\% ± 2.80\%  \\ 
\hline
\end{tabular}
\end{table}

\vspace{5pt} 

\begin{table}[h]
\centering
\caption{Relative / Abs error rate after complement per step.}
\label{Tab:complement_error}
\renewcommand{\arraystretch}{1.2} 
\setlength{\tabcolsep}{8pt} 

\begin{tabular}{c|c} 
\hline
Location & Error rate $\pm$ Standard deviation \\ 
\hline
Inside Lab (Fig.\ref{fig:testbed})  
& -0.0248\% ± 0.0981\% / 0.0744\% ± 0.0686\%  \\ 
\hline
Room to room  
& -0.213\% ± 0.400\% / 0.355\% ± 0.314\%  \\ 
\hline
\end{tabular}
\end{table}

\section{Evaluation}
\subsection{Environmental Setup and Data Collection}
\color{black}

The experiments were held in our lab (big cluttered environment\footnote{The environment includes many objects
and furniture that may cause blocking or attenuation of the light beam.}) that spans an area of $18m\times11m$. The room is illuminated with $30$ led light sources at a height $2.7m$ from the floor at equal intervals. 

Figure~\ref{fig:testbed} provides an overview of the experimental testbed, highlighting the predefined walking paths, the placement of obstacles, and the lighting conditions. The layout was designed to introduce variations in light intensity due to furniture and other environmental factors, simulating real-world conditions for photocurrent-based sensing. The predefined paths included a mix of straight-line segments and 90-degree turns, ensuring diverse movement patterns that challenge the robustness of the proposed system. These routes were recorded using an annotation application for precise tracking and analysis.
\color{black}

All the experiments were conducted using \textit{four} solar panels attached to the shoes worn by the considered participants. 
Two solar panels were attached one is on the side and another on the top center of each shoe of the pair (i.e., the left shoe and the right shoe), as shown in Fig~\ref{fig:demo_shoes}.
The sampling rate was set to 28Hz for all cells. 
\textit{Six} different participants of different genders and body shapes (e.g., fat and thin) were involved in the experiments and the data collection process.

Each participant walked five laps along the predefined routes, following their natural walking patterns. Photocurrent measurements were recorded in real-time from all four solar cells.
\color{black}


\subsection{Evaluation of Context Recognition}
\subsubsection{Step Counting Performance}
Figure \ref{fig:confusion_matrix} shows the confusion matrix of the proposed step counting system. 
The real implementation of the system confirms the high accuracy of estimating the correct step counts of the four users with $88\%$  of the time. Additionally, if we allow two-steps error, the system achieves a perfect step count accuracy of 100\%. This is achieved while the measuring device (solar cell) performs the required sensing without consuming energy but instead harvesting energy.  

 Table \ref{Tab:step_detector_error} and \ref{Tab:complement_error} details the error rate and standard deviation per step for transfers within a particular room and between different rooms, based on a sequence of approximately 1000 steps per user. Additionally, the mean relative percentage error and mean absolute percentage error, along with the standard deviation of each error, were calculated for these sequences.
\color{black}

The photovoltage signal depicted in Fig. \ref{fig:loc_photocurrent} resulted from variations in the lighting environment as the subject navigated the room. Initially, the error rate in step estimation escalated to 5.9\%; however, it was effectively reduced to 0.35\% through the implementation of the missing step complement module.
We believe this result opens the door for the next generation of battery-less context-aware sensing systems.

\subsubsection{Subject Identification Performance}
\begin{figure}[t]
\begin{center}
\includegraphics[width=.7\linewidth]{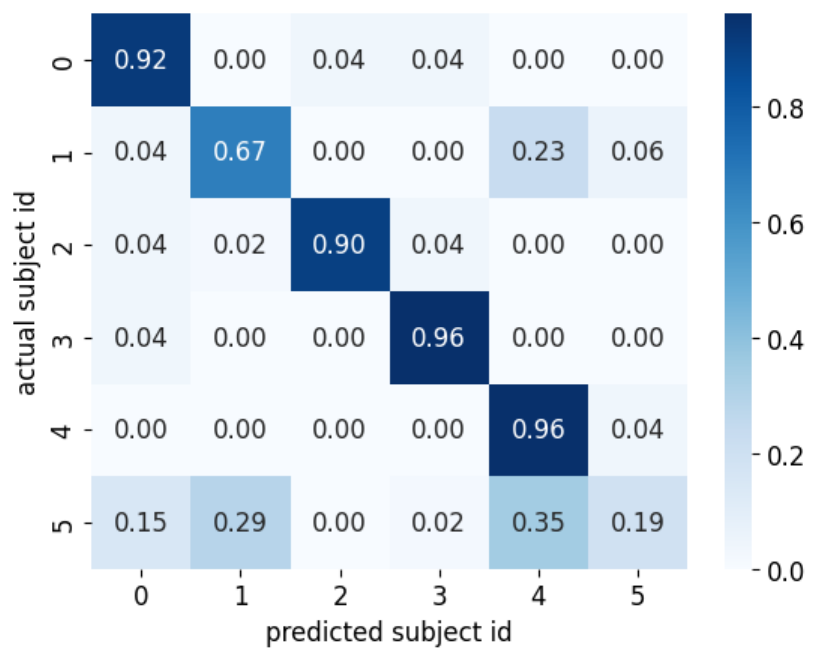}
\caption{
   The confusion matrix of subject identification using KNN-based Euclidean similarity
}
\label{fig:cm_noDTW}
\end{center}
\end{figure}

\begin{figure}[t]
\begin{center}
\includegraphics[width=.7\linewidth]{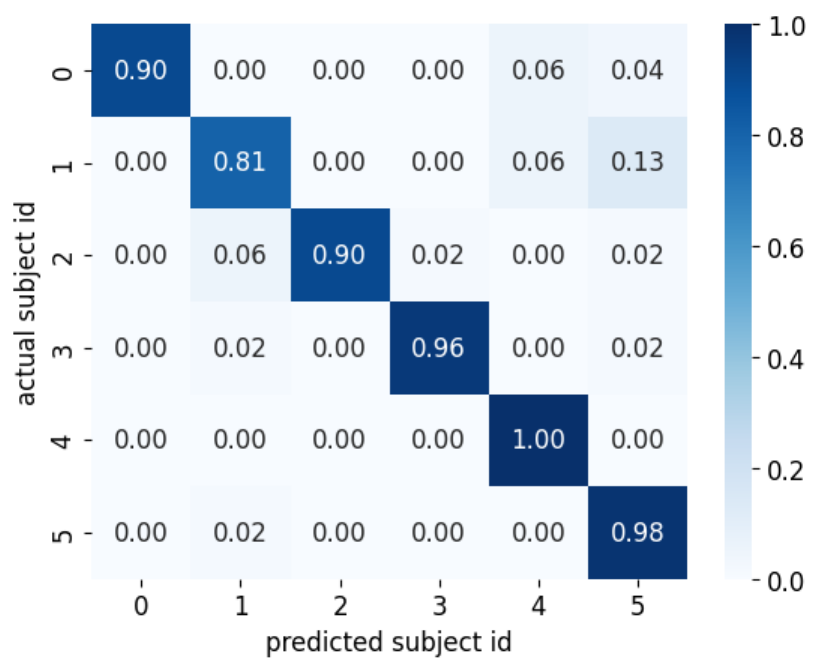}
\caption{
   The confusion matrix of subject identification using the proposed KNN-based modified DTW similarity
}
\label{fig:cm_DTW}
\end{center}
\end{figure}

In this section, we evaluate the system's ability to recognize the subject's identity from his gait pattern.
Figure~\ref{fig:cm_DTW} shows that the system can accurately differentiate  between different subjects with an accuracy of 92.5\%. This result can be justified due to the ability of the proposed modified DTW to characterize the mobility pattern of each user given  the corresponding photocurrent measurements. This is better than the traditional matching with Euclidian similarity, shown in Figure~\ref{fig:cm_noDTW}, by 16\%.
This confirms the superiority and flexibility of the  matching based on the modified DTW in recognizing the user step pattern even with varying speeds (i.e., stretched or shrunk step pattern) compared to the  kNN-based  Euclidian.
This highlights the system's recognition ability as energy-free sensing for gait and subject recognition.

\subsubsection{Localization Performance}

\begin{figure}[t]
\begin{center}
\includegraphics[width=0.8\linewidth]{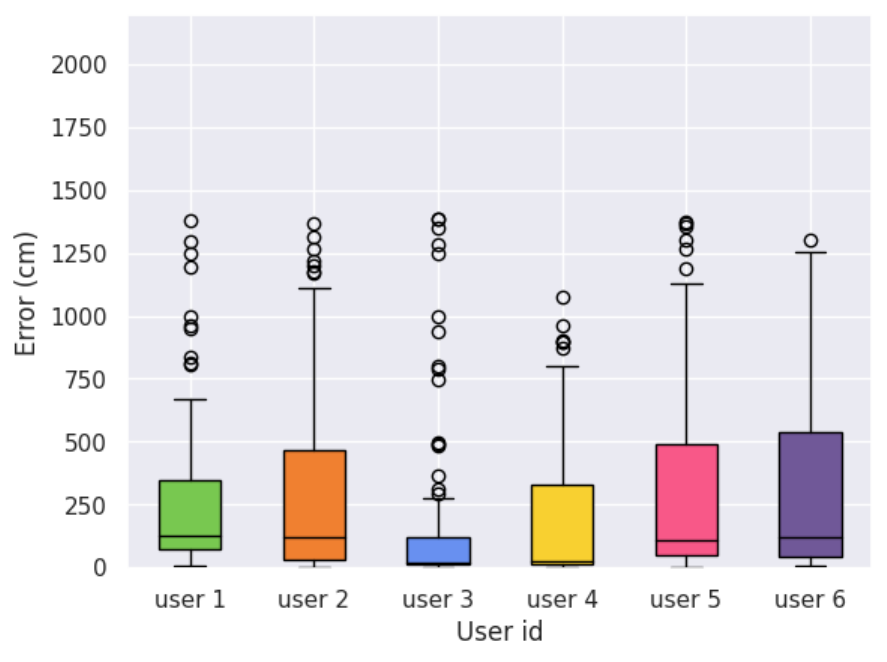}
\caption{
    Localization of each user
}
\label{fig:each_user}
\end{center}
\end{figure}

In this section, we evaluate the proposed system's ability to accurately localize different subjects in a realistic environment. To assess the performance of the localization algorithm, we employed a leave-one-out evaluation method, in which the estimator has access to all data except that of the user currently being tested. The results, as shown in Figure~\ref{fig:each_user}, are presented in the form of boxplots, which display the localization error of different users as computed by the Euclidian distance between the true location and the estimated location. The results indicate that the system is able to accurately track the location of different subjects, with a median location error of 43cm, indicating a high degree of precision. This promising outcome can be attributed to the system's ability to recognize the light intensity at each location in the area of interest, allowing it to accurately estimate the user's location. Furthermore, the results demonstrate that the system's localization performance is robust to variations in the mobility patterns of different users, with a maximum error of around 1m, further highlighting its potential as a reliable energy-free sensing technology for context recognition and localization.

\begin{figure}[t]
\begin{center}
\includegraphics[width=0.75\linewidth]{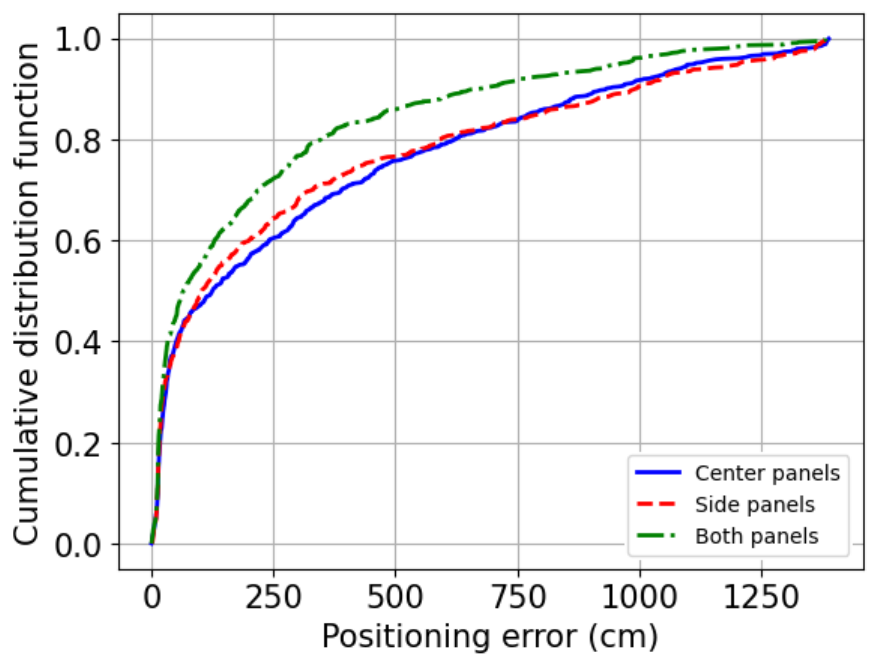}
\caption{
   The comparison between side and center and both panels
}
\label{fig:side_center_both_cdf_5sec}
\end{center}
\end{figure}

\begin{figure}[t]
\begin{center}
\includegraphics[width=0.8\linewidth]{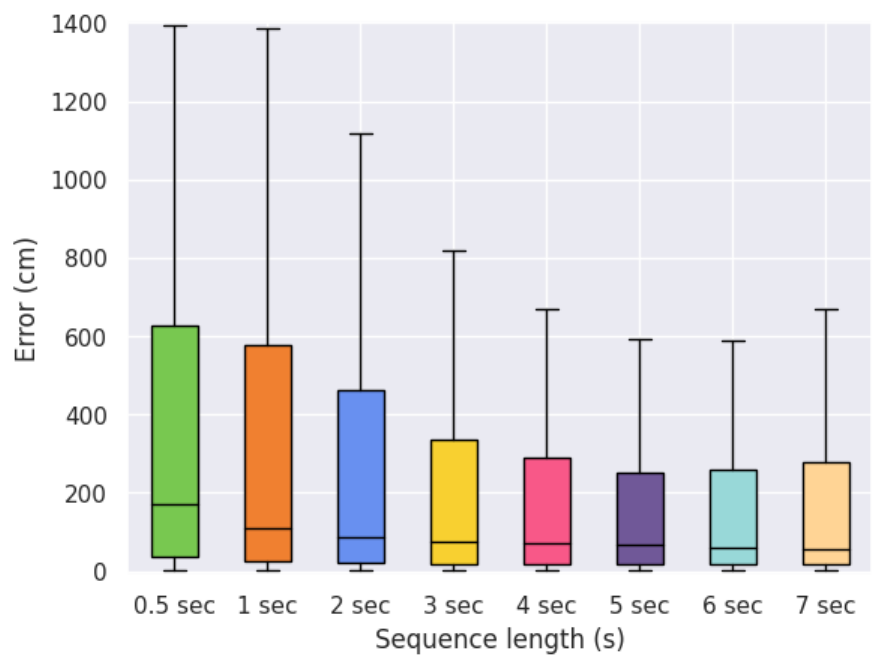}
\caption{
   The effect of changing sequence length
}
\label{fig:error_box_plot_diff_time}
\end{center}
\end{figure}

\subsection{Effect of System Parameters}

\subsubsection{Position of Solar Cell}
In this section, we explore the impact of the placement of solar cells on the localization performance of the system. We attached two cells to each shoe, one placed at the top center and another on the outer side, to analyze their individual and combined effects.
The results, as shown in Figure ~\ref{fig:side_center_both_cdf_5sec}, are represented in the form of cumulative distribution function (CDF) of the positioning error. The figure highlights the fact that the top-center cell pair outperforms the outer-side cell pair in terms of localization accuracy. This is likely due to the direct exposure of the top-center cells to the light source, resulting in a clearer and less ambiguous location pattern.
Moreover, the figure also shows that the use of both cell pairs simultaneously leads to the best performance. This is because the system receives more information (more measurements from different exposures) about the location from the combined cells, providing a more comprehensive and accurate picture of the context.

\subsubsection{The effect of the similarity method}
\begin{figure}[!t]
\begin{center}
\includegraphics[width=0.8\linewidth]{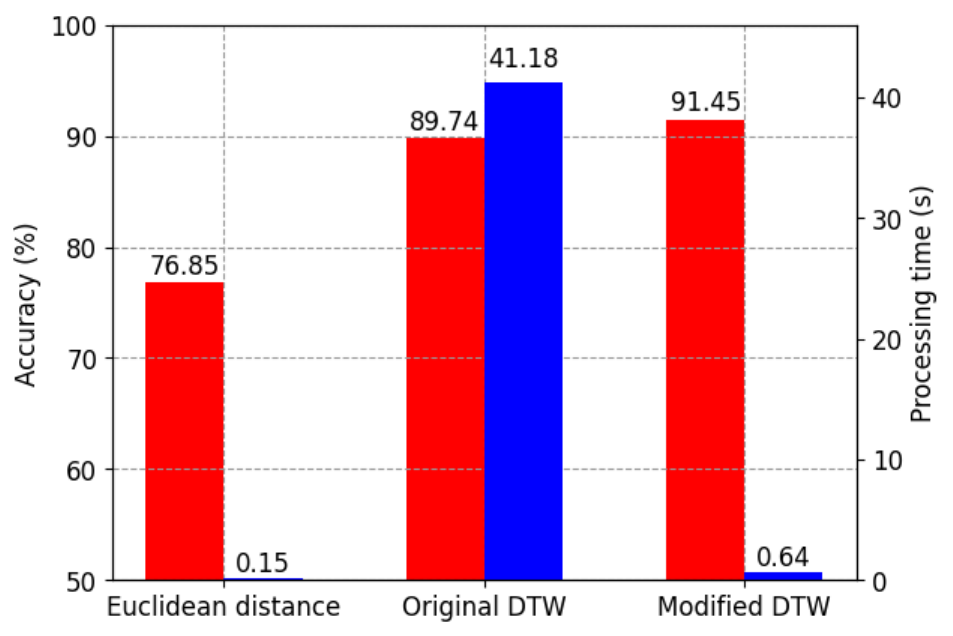}
\caption{
   The subject identification performance of different similarity measures.
}
\label{fig:acc_time}
\end{center}
\vspace{-0.5cm}
\end{figure}

In this section, we assess the effectiveness of our proposed subject identification technique that utilizes the modified DTW algorithm. In Figure~\ref{fig:acc_time}, we compare this method with the traditional DTW and Euclidean distance similarity algorithms. The results demonstrate the superiority of our proposed method with an accuracy of 91.5\% and a processing time of 0.64 seconds. On the other hand, the traditional DTW algorithm achieved an accuracy of 89.74\% with a processing time of 41.18 seconds, while the Euclidean distance similarity method achieved an accuracy of 76.85\% with a processing time of 0.15 seconds.
The results indicate that our proposed method provides a better balance between accuracy and processing time compared to the other methods. Our method achieved a significant improvement in accuracy, with an increase of 5.76\% and 14.65\% compared to the traditional DTW and Euclidean distance methods, respectively. This is done while maintaining a fast processing time that meets real-time requirements. These findings highlight the potential of our modified DTW algorithm as an efficient and effective method for subject identification.

\subsubsection{The effect of the sequence length}
Here, we examine the impact of the sequence length used by the modified DTW algorithm on the localization error. The results, shown in Figure~\ref{fig:error_box_plot_diff_time}, demonstrate the relationship between the sequence length and the localization error. Initially, a sequence length of 0.5 seconds results in high localization error, but as the sequence length increases, the error decreases, reaching its minimum value at a sequence length of 5 seconds.

This minimum value represents the best balance between accuracy and robustness, as the sequence length of 5 seconds provides sufficient information about the user's location while still being able to capture their movement effectively. However, beyond a certain point, the error starts to increase again, indicating that longer sequences may contain information from multiple locations and lead to a reduced accuracy.
These results highlight the importance of selecting an appropriate sequence length for the modified DTW algorithm to achieve optimal localization accuracy. The findings of this evaluation also support the suitability of the kNN-based modified DTW  as a valuable tool for enabling real-time location tracking.

\subsubsection{Computational time}

In this section, we evaluate the real-time performance of the system.  The step-counting algorithm detects a step in less than 1ms. This is due to the relatively simple computations performed by the peak estimation, which allows the algorithm to run on micro-computing units such as microcontrollers and microbits attached to the shoes. Additionally, this minimal computing time is significantly faster than the average step time of 0.5 seconds for humans.
The results confirm the responsiveness of the system, enabling its real-time operation. 
Additionally, Figure~\ref{fig:acc_time} illustrates the performance of subject identification, comparing accuracy and processing time across three similarity measures: Euclidean distance, DTW, and the proposed modified DTW. The results indicate that the proposed method achieved an accuracy of 91.45\% while requiring only 0.64 seconds of processing time on a single thread. This represents a substantial improvement over the traditional DTW method in terms of both efficiency and effectiveness. 
However, while the Euclidean distance method exhibits a marginally faster processing time, it incurs a significant decline in performance. This discrepancy can be attributed to the inherent characteristics of the Euclidean distance measure, which calculates the straight-line distance between two points in space. This approach, while computationally less intensive, often fails to capture the dynamic temporal variations in time-series data, leading to reduced accuracy in applications like subject identification where temporal patterns are crucial.
These findings underscore the effectiveness of the modified DTW method, which enhances the classic DTW algorithm by optimizing the path through the time series data and reducing unnecessary computations. This leads to a balanced trade-off between processing time and accuracy, desirable for the proposed system.

\begin{table}[!t]
    \centering
    \caption{Power consumption details of the proposed system with QN9080 chip.}
    \begin{tabular}{|c|c|}
       \hline
       \textbf{System Part} & \textbf{Power Consumption} \\
       \hline
       ADC  & 0.42 mW   \\
       \hline
       Communication (constant)   & 10.5 mW  \\
       \hline
        Communication (intermittent)   & 0.7  mW  \\
       \hline
       Calculation & 1.44 - 1.77 mW  \\
       \hline \hline
       Total (intermittent) & 2.56 - 2.89 mW \\
       \hline
    \end{tabular}
    \label{tab:power_consumption_table}
\end{table}

\begin{figure}[!t]
    \centering    \includegraphics[keepaspectratio, scale=0.2]{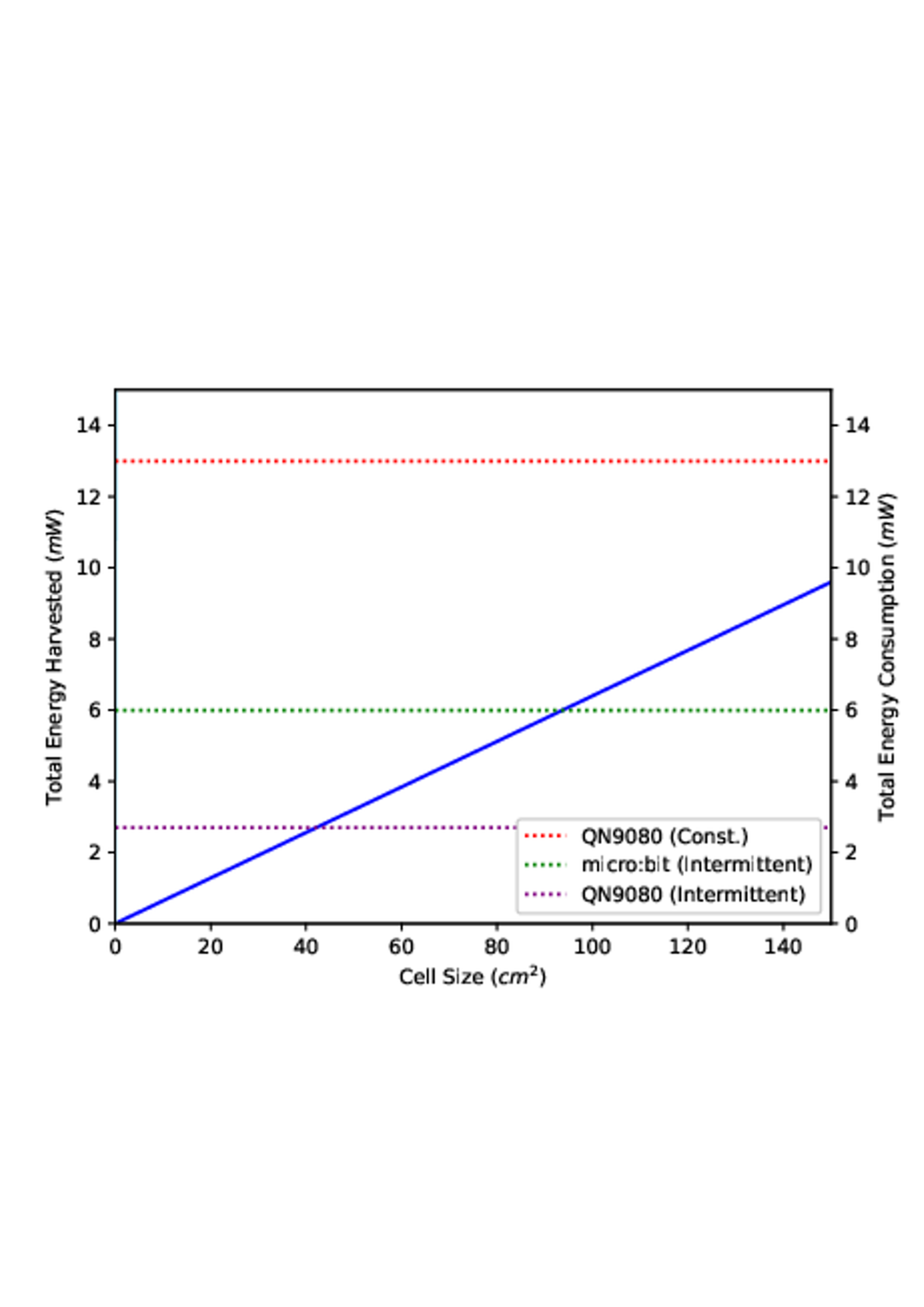}
    \caption{The effect of solar cell size on the harvested energy and comparing it to and demanded power for the system operation.}
    \label{fig:har_con}
\end{figure}

\subsubsection{Energy Consumption}\label{sec:eval_energy_consumtion}
In this section, we examined how much energy is used by our system for processing signals, performing computations, and transferring data. 
In our implementation, we used a Micro:bit device due to its ease of use in our experiments. However, it consumes unnecessary energy because its lights remain on, even when it is idle. Thus, for more efficient implementation, we leveraged the QN9080 chip as it has a lower power consumption profile. Our findings, shown in Table \ref{tab:power_consumption_table}, summarize the power consumption of different parts of the system. 
The process of converting the light signal (photo-voltage) into digital data that the computer can process also consumes power. Specifically, the power usage of the analog-to-digital converter in the QN9080 chip ranges from 200 to 350 microamperes. Additionally, the computation power involves the power necessary for the CPU to read sensor data and run our algorithm. The table also indicates that the BLE communication part uses more energy compared to other parts.
To manage this high energy consumption, the system employs an intermittent communication strategy where the system only transfers data once every hour, significantly reducing energy usage. 
Specifically, the system transfers 86.4 Kbytes of data at 3 Hz which takes about 5 seconds over BLE, leading to a total power consumption of approximately 2.56 to 2.89 milliwatts, as detailed in Table \ref{tab:power_consumption_table}.

Finally, Figure \ref{fig:har_con}  compares power harvesting and consumption, showing how the solar cell size affects the system's energy requirements. Taking into account that the efficiency of the solar panels can decrease when they are tilted—which can happen with user movement—we estimated that the panels operate at 60 microwatts per square centimeter on average, with a 19\% power conversion efficiency\footnote{The harvested energy in an indoor environment (illuminance: $1000 lux$, light intensity $0.31 
 mW/cm^2$) is maximum 77.5 \textmu$W/cm^2$ at 25\% PCE(Power Conversion Efficiency) ~\cite{Penpong2023}.}. Given these parameters, a solar panel size of approximately $50 cm^2$ would suffice for our system’s power needs. Since the available surface area on a shoe, excluding the bottom and inside, generally exceeds $500 cm^2$, it's plausible to accommodate the solar-powered system effectively. This conclusion underscores the practicality of integrating our energy-efficient system into everyday objects, leveraging solar power to meet operational energy requirements.

\section{Related Work} \label{sec:related_work}

\subsection{Step Count Technologies}
The evolution of step count technologies has garnered considerable attention in the technology and health sectors over recent years, reflecting a growing interest in monitoring and promoting physical activity.

\begin{table*}[htbp]
    \centering
    \caption{Comparison of Different Step Counting and Identification Methods}
    \renewcommand{\arraystretch}{1.3}
    \begin{tabular}{p{2.1cm} p{2cm} p{1.3cm} p{1.2cm} p{1.5cm} p{1.5cm}}
        \toprule
        \textbf{Method} & \textbf{Sensing Mechanism} & \textbf{Energy Harvesting} & \textbf{Step Count} & \textbf{User Identification} & \textbf{Localization} \\
        \midrule
        IMU-based \cite{6974989, 7885036} & \raggedright Accelerometer & No & 99\% & 89\% & drift error \\ \hline
        WiFi CSI \cite{ 10.1145/3084041.3084061} & \raggedright RF Reflections & No & N/A & 80-90\% & 1-1.5m error \\\hline
        RFID-based \cite{10.1145/2639108.2639111} & \raggedright Tag-based Tracking & No & N/A & 85\% & 50cm error \\ \hline
        Camera-based (Gait) \cite{gait_recognition, Khaliluzzaman2023-xk} & \raggedright Vision-based (RGB) & No & N/A & 99\% & $<$5cm error \\\hline

        Pneumatic Energy Harvesting \cite{Shveda2022-qp} & \raggedright Air Pressure & Yes & N/A & N/A & N/A \\
\hline
        Piezoelectric-based \cite{928763} & \raggedright Pressure-based  & Yes & 89\% & N/A & N/A \\\hline
        Optical (Ours)  & \raggedright Light intensity variation & Yes & 89\% & 90\% & 43cm error \\
        \bottomrule    
    \end{tabular}
    \color{black}

    \label{tab:comparison}
\end{table*}

As summarized in Table~\ref{tab:comparison}, several step counting technologies exist, each with its advantages and limitations. 

\color{black}

A predominant method in this domain is accelerometer-based step counting, which has proven to be an effective means of tracking physical activity. This technique utilizes accelerometer sensors integrated into various wearable and portable devices, including smartphones \cite{6974989, naqvib2012step, 7885036}, smartwatches \cite{7921535}, and footwear \cite{8036783}.

These sensors capture the acceleration of the user's body movements, enabling the computation of steps taken through sophisticated data processing algorithms.
Moreover, innovative developments have led to the integration of pedometers directly into footwear, powered by energy-harvesting technologies. Notable among these are designs that feature ferroelectric insoles, which not only generate power from mechanical stress but also serve as event detection mechanisms \cite{10.1109/SAS.2017.7894114}. Additionally, shoe-mounted piezoelectric devices have been explored for their dual functionality in energy scavenging, and step counting \cite{928763,7511782}. These advancements highlight a trend toward self-powered and multifunctional wearable devices.
However, despite the advancements in step-counting technology, certain limitations persist. Specifically, the current generation of shoe-integrated systems primarily focuses on step quantification and lacks the capability for advanced user identification or location tracking. Furthermore, the long-term durability and reliability of these integrated sensors remain underexplored, necessitating further research to ensure their viability in everyday use.

\subsection{User Identification Technologies}
User identification technologies have seen substantial advancement, with a variety of methods developed to recognize individuals under different conditions.
Conventional techniques utilize RGB cameras, which are effective in well-lit environments and have been reported to achieve an accuracy rate of around 80\% \cite{9714177, gait_recognition}. In contrast, for dimly lit settings, low-power radar \cite{Vandersmissen2018IndoorPI} and Micro-size LiDAR systems \cite{yamada2022accurate} have been employed, boasting higher accuracy rates of above 95\% and 96\%, respectively.
An innovative approach known as FootprintID \cite{10.1145/3130954} employs a passive vibration sensor to analyze walking patterns, utilizing an Incremental Time Support Vector Machine (ITSVM) for classification and achieving a noteworthy identification success rate of approximately 90\%.
Wi-Fi Channel State Information (Wi-Fi CSI) has also been harnessed to authenticate users in indoor environments, demonstrating an accuracy range of 80\%-90\% \cite{10.1145/3084041.3084061,10.1145/2971648.2971670,zeng2016wiwho}. Remarkably, CSI-Net \cite{zeng2016wiwho} extends the capabilities of Wi-Fi CSI by incorporating Convolutional Neural Networks (CNNs) to enhance user recognition alongside biometric analysis, gesture recognition, and fall detection, illustrating the multifaceted potential of these networks.

Despite these advancements, their performance can be compromised by interference in cluttered environments. Moreover, there are unresolved issues related to the high dependency on existing infrastructure and the economic burden associated with deploying and maintaining these sensor systems. Further research is necessary to address these limitations and improve the viability of user identification technologies in various settings.

\subsection{Energy-Efficient Localization Technologies}
Indoor location estimation technologies are vital in today's interconnected world, with a significant focus on developing methods that are not only accurate but also cost-effective and energy-efficient \cite{Zafari2017ASO, 10.1145/2933232, 6407455}.
Wireless sensing technologies play a pivotal role in this area, employing a variety of methods such as received signal strength (RSS) \cite{rizk2018cellindeep,abbaswideep}, time of arrival (ToA)~\cite{mohsen2023locfree}, angle of arrival (AoA), and time difference of arrival (TDoA)~\cite{mohsen2023privacy}. Additionally, hybrid approaches combine these techniques to enhance accuracy and reliability \cite{Zafari2017ASO,youssef2022magttloc,rizk2022robust}.
The integration of backscatter communication technology significantly enhances indoor localization accuracy while emphasizing energy efficiency. By employing backscatter tags, which operate by reflecting and modulating existing radio frequency (RF) signals instead of generating new ones, location estimation achieves a median error remarkably low, ranging from 0.9 to 1.5 meters in environments equipped with two or more access points (APs) \cite{10.1145/3143361.3143379}. Similarly, Slocalization \cite{10.1109/IPSN.2018.00052} operates on sub-microwatt power consumption levels while eliminating power-intensive components, such as WiFi transmitters, from the tags. However, challenges like signal interference and wave attenuation become more pronounced as the number of devices in an environment increases.
Other studies \cite{zigbee1, uradzinski2017advanced}  utilize Zigbee wireless technology, which is valued for its low power consumption profile compared to traditional wireless technologies. These studies exploit the strengths of Zigbee by using its reliable signal characteristics and energy-efficient operation to build and utilize fingerprint databases for improved localization accuracy, employing matching algorithms such as the Nearest Neighbor \cite{zigbee1} and Bayesian algorithm \cite{uradzinski2017advanced}.
Nevertheless, power consumption remains a concern, with typical values around 10 milliwatts, indicating a need for further optimization in this technology segment.

On the other hand, Pedestrian Dead Reckoning (PDR) is particularly effective when sensors are firmly attached to a shoe, with a 3D accelerometer consuming approximately 0.32 mW \cite{320micro}. Although theoretically capable of estimating relative position by double-integrating acceleration data, PDR faces challenges with accumulated drift errors over time, making long-term location tracking difficult without corrections \cite{6699780}. Common strategies include combining accelerometer data with gyro sensor inputs to estimate step count or stride length and detect body orientation, though these do not fully mitigate error accumulation inherent in PDR systems.

\subsection{Light-based Sensing and Energy Considerations}
The integration of solar panels in light-based sensing devices has emerged as a promising avenue for energy-efficient applications, including gesture recognition \cite{ma2018gesture, SolarGest_simulator}, indoor location tracking \cite{9767256,randall2007luxtrace,umetsu2019ehaas,10.5555/3578948.3578961}, and global positioning \cite{sunspot}. These systems not only serve their primary function but also harness solar energy, contributing to their energy sustainability.
LuxTrace \cite{randall2007luxtrace} is an exemplar of this integration, where solar cells mounted on a user's shoulder serve dual purposes: capturing light intensity for precise location tracking, with an accuracy of 21 cm at the 80th percentile, and converting this light into usable energy.
Another noteworthy system is ZEL \cite{9762376}, a net-zero-energy lifelogging tool designed for office environments. It uses a wearable device with heterogeneous energy harvesters, achieving remarkable energy efficiency of 99.6\% and recognition accuracy of 93.1\% of places and activities.
Solar panels, sized at 48 $cm^2$ and worn on the wrists and feet, have shown the innovative potential to classify seven different user activities with 91.7\% accuracy \cite{10.1145/3594739.3610744}. These applications underscore the dual functionality of solar panels as both sensors and energy harvesters.
PVDeepLoc \cite{9767256}, a passive solar-panel sensor system, shows promising results with median localization errors under 75 cm. However, real-world applications, especially in complex indoor settings with furniture, are yet to be fully explored.

\textit{Contrasting with existing solutions, our proposed system aims to pioneer the use of solar panels on footwear. This novel approach is designed to accurately estimate step counts, identify users, and pinpoint their locations, all while maintaining a net-zero energy balance for both sensing and computation. This leverages the renewable energy from the ambient light, powering the device and optimizing the necessary computations.}

\section{CONCLUSION}
In conclusion, this paper introduces a novel energy-free sensing system that leverages photovoltaic cells for step counting, localization, and human identification. Utilizing foot-mounted photovoltaic cells, this approach capitalizes on their dual functionality as sensors for detecting steps, user identity, and location, as well as sustainable energy sources for powering the computing unit. The system operates effectively under various environmental and lighting conditions.
The photovoltaic cells capture distinct mobility patterns essential for user identification, step counting, and localization tasks. Through the application of computationally efficient Dynamic Time Warping (DTW), our system aligns photocurrent readings with pre-existing patterns, accurately determining the user's identity and location. The evaluation conducted in realistic settings validates the system's capabilities, demonstrating promising results with accuracies of 88\% for step counting, 90\% for user identification, and a localization precision within 43 cm.
Our research highlights the potential of photovoltaic technology in creating non-intrusive, self-sustaining sensing systems that operate independently of external power sources. This advancement fosters broader adoption and development in the field of energy-efficient sensing solutions, signifying a substantial step forward in integrating sustainable technologies within human-centric applications.

\textbf{ACKNOWLEDGMENTS}
This work was supported in part by JSPS KAKENHI Grant number 22K12011 and 
19H05665, Japan.

\bibliographystyle{unsrt}

\end{document}